\documentclass{article}

\usepackage{arxiv}
\usepackage[utf8]{inputenc}
\usepackage[T1]{fontenc}
\usepackage{hyperref}
\usepackage{url}
\usepackage{booktabs}
\usepackage{amsfonts}
\usepackage{amsmath}
\usepackage{nicefrac}
\usepackage{microtype}
\usepackage{lipsum}
\usepackage{graphicx}
\usepackage[numbers]{natbib}
\usepackage{doi}
\usepackage{cleveref}
\usepackage{amsthm}
\usepackage{xcolor}
\usepackage{soul}
\usepackage{bbm}
\usepackage[bbgreekl]{mathbbol}
\usepackage{rotating}

\newcommand{\CenteredVcell}[2]{\begin{centering}\begin{sideways}\parbox[c]{#1}{\begin{centering}{\hfill #2 \hfill\textcolor{white}{.}}\end{centering}}\end{sideways}\end{centering}}

\newcommand{\setC}{\mathcal{C}}
\newcommand{\setI}{\mathcal{I}}

\usepackage{todonotes}

\usepackage{algorithm}
\usepackage{algpseudocode}

\newtheorem*{remark}{Remark}

\DeclareMathOperator*{\argmin}{argmin}

\DeclareMathOperator{\rank}{rank}

\title{Sampling Strategies in Bayesian Inversion: A Study of RTO and Langevin Methods\thanks{This work was funded by a Villum Investigator grant (no. 25893) from the Villum
Foundation.}}

\author{ \href{https://orcid.org/0009-0008-8364-7761}{\includegraphics[scale=0.06]{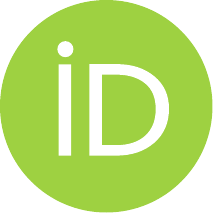}\hspace{1mm}Rémi Laumont \thanks{Corresponding author}} \\
	Department of Applied Mathematics and Computer Science\\
	Technical University of Denmark\\
	Kongens Lyngby, 2800 \\
	\texttt{real@dtu.dk} \\
	\And
	\href{https://orcid.org/0000-0001-8363-9448}{\includegraphics[scale=0.06]{orcid.pdf}\hspace{1mm}Yiqiu Dong} \\
	Department of Applied Mathematics and Computer Science\\
	Technical University of Denmark\\
	Kongens Lyngby, 2800 \\
	\texttt{yido@dtu.dk} \\
    \And
	\href{https://orcid.org/0000-0002-4654-3946}{\includegraphics[scale=0.06]{orcid.pdf}\hspace{1mm}Martin Skovgaard Andersen} \\
	Department of Applied Mathematics and Computer Science\\
	Technical University of Denmark\\
	Kongens Lyngby, 2800 \\
	\texttt{mskan@dtu.dk} \\
}

\date{}

\renewcommand{\headeright}{}
\renewcommand{\undertitle}{}
\renewcommand{\shorttitle}{Sampling Strategies in Bayesian Inversion}

\hypersetup{
	hidelinks,
pdftitle={A template for the arxiv style},
pdfsubject={q-bio.NC, q-bio.QM},
pdfauthor={Rémi Laumont, Yiqiu Dong, Martin S. Andersen},
pdfkeywords={Inverse problems, sampling, RTO, Langevin methods, deblurring, inpainting, parameter selection},
}

\begin{document}
\maketitle

\begin{abstract}
	This paper studies two classes of sampling methods for the solution of inverse
	problems, namely Randomize-Then-Optimize (RTO), which is rooted in sensitivity
	analysis, and Langevin methods, which are rooted in the Bayesian framework.
	The two classes of methods correspond to different assumptions and yield
	samples from different target distributions. We highlight the main conceptual
	and theoretical differences between the two approaches and compare them from
	a practical point of view by tackling two classical inverse problems in imaging:
	deblurring and inpainting. We show that the choice of the sampling method has
	a significant impact on the reconstruction and the proposed uncertainty.
\end{abstract}



\section{Introduction}\label{sec:intro}

A typical inverse problem in imaging is to retrieve an image $x \in \mathbb{R}^d$ from a
degraded observation $y \in \mathbb{R}^m$. A common observation model is the additive noise
model, which can be formulated as
\begin{equation}\label{eq:inv_pb}
	y = \mathcal{A}(x) + n, 
\end{equation}
where $\mathcal{A} \colon \mathbb{R}^d \to \mathbb{R}^m$ is a so-called forward operator that
models the deterministic aspects of the observation process. The term $n$ represents additive
noise, and we will henceforth assume that $n$ is zero-mean Gaussian white noise with covariance
$\sigma^2 I$. We will also assume that the forward operator $\mathcal{A}$ is known and linear,
which allows us to express the measurement model as $y = Ax + n$ with
$A\in\mathbb{R}^{m\times d}$. 

The problem of estimating $x$ from the observation vector $y$ is often ill-posed. Generally
speaking, this means that the observation may not correspond to a unique reconstruction or that
the reconstruction is very sensitive to perturbations of $y$. Regularization in the form of
prior information on $x$ is then necessary to obtain a well-posed problem
\cite{hansen2021computed}. In the variational framework, the reconstruction problem takes the
form of an optimization problem,
\begin{equation}\label{eq:variational_framework}
	x^* \in \argmin_{x} \, \{ f(Ax, y) + g(x) \},
\end{equation}
where $x^*$ denotes the reconstruction. The objective function consists of a data-fidelity term
$f(Ax,y)$ and a regularization term $g(x)$. The data-fidelity term measures the discrepancy
between the observation $y$ and the model output $Ax$ whereas the regularization term $g(x)$
penalizes images with undesirable properties, e.g., images outside a neighborhood of some
sub-manifold. In many cases, the minimization problem in \eqref{eq:variational_framework} can be
solved efficiently using advanced optimization methods. However, a solution to the problem in
\eqref{eq:variational_framework} does not provide information about the inherent uncertainty
that may arise because of noisy measurements, discretization, and/or model errors. 

Information about uncertainty can be obtained by adopting a probabilistic approach, and the
Bayesian framework is a natural choice for this purpose. Assuming that the true image $x$ and
the observation $y$ are considered realizations of random variables, the posterior probability
density function (pdf) of $x$ given $y$ can be expressed in terms of the pdf $\pi_{\mathbbm{y}|\mathbbm{x}=x}(y)$, which
characterizes the observation model, and a prior pdf $\pi_{\mathbbm{x}}(x)$, which encodes any prior
information we may have about $x$ before observing $y$. Using Bayes' formula, we can express 
the posterior density as
\begin{equation}
	\pi_{\mathbbm{x}|\mathbbm{y}=y}(x) = \frac{\pi_{\mathbbm{x}}(x) \  \pi_{\mathbbm{y}|\mathbbm{x}=x}(y)}{\pi_{\mathbbm{y}}(y)}, \qquad  \pi_{\mathbbm{y}}(y) = \int_{\mathbb{R}^d} \pi_\mathbbm{x}(x)\  \pi_{\mathbbm{y}|\mathbbm{x}=x}(y)\, \mathrm{d}x.
\end{equation}   
From a Bayesian point of view, the posterior $\pi_{\mathbbm{x}|\mathbbm{y}=y}(x)$ is the solution
to the inverse problem since it provides a complete characterization of the uncertainty. The
posterior density can be used to compute point estimates, such as a maximum a posteriori (MAP)
estimate or the posterior mean, and to quantify uncertainty in the form of credible intervals
or second-order moments.

Although the Bayesian framework is well-established from a theoretical point of view, the
posterior density is often intractable and cannot be computed in closed form. Thus, the
posterior density is often explored using sampling methods. These generate a set of samples
$\{x^{(i)}\}_{i=1}^N$ from the posterior density $\pi_{\mathbbm{x}|\mathbbm{y}=y}(x)$,
allowing us to compute point estimates and quantify uncertainty using Monte Carlo
approximations to high-dimensional integrals. 

\begin{figure}[htbp]
	\centering
	\includegraphics[width=0.32\textwidth]{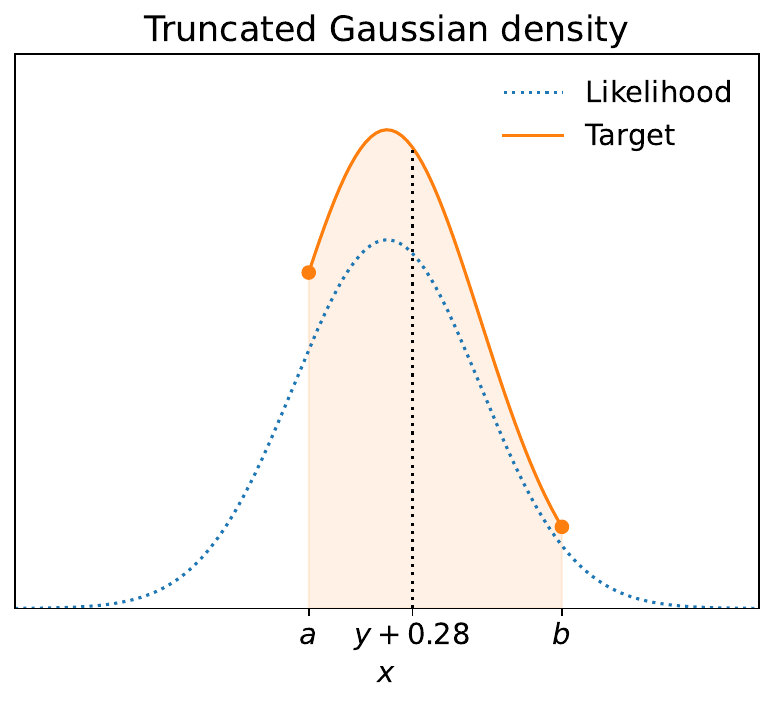}\ 
	\includegraphics[width=0.32\textwidth]{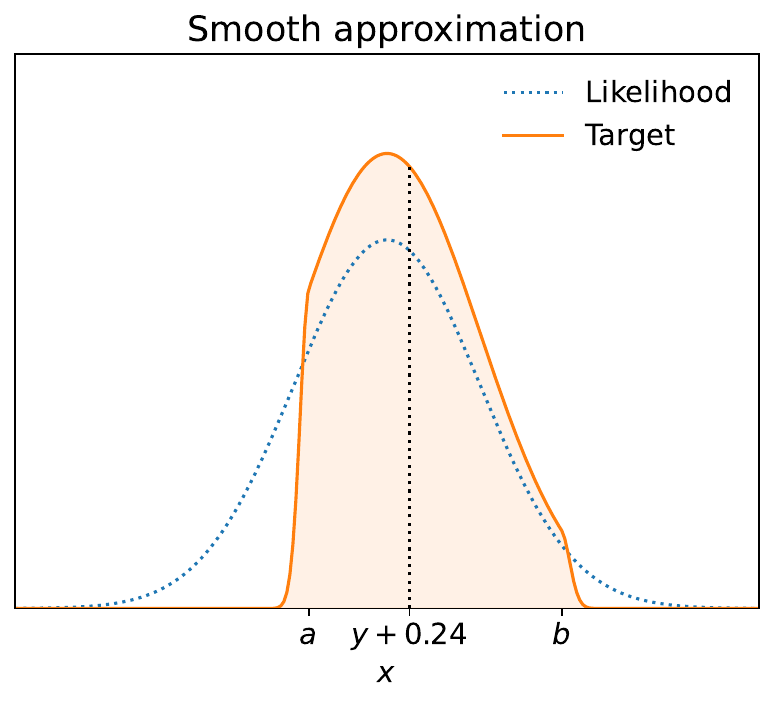}\ 
	\includegraphics[width=0.32\textwidth]{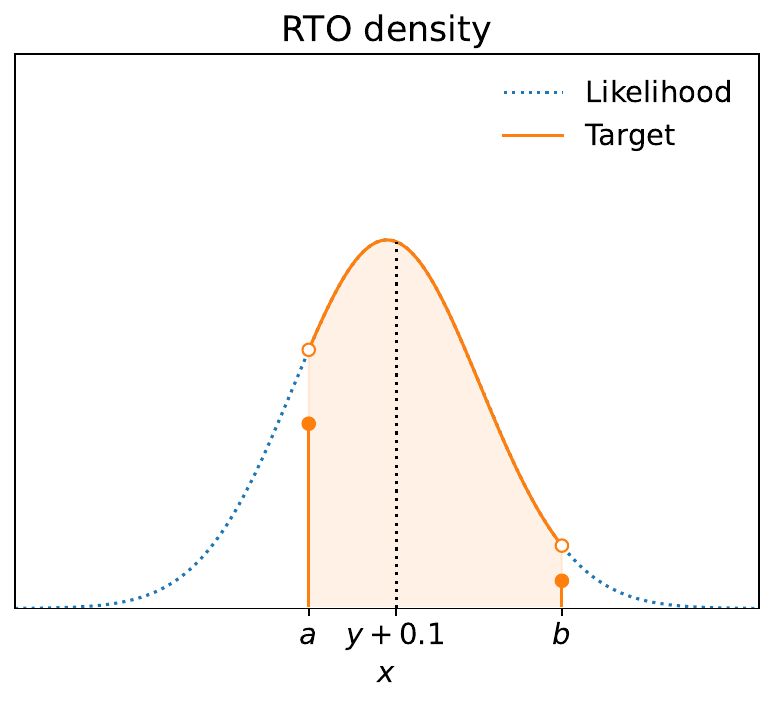}
	\caption{Truncated Gaussian posterior, a smooth approximation, and the RTO density for a
	Gaussian observation model. The vertical dotted line marks the target mean relative to
	the observation $y$, which is the mode of the likelihood.}\label{fig:posterior_comparison}
\end{figure}

The purpose of this paper is to contrast and compare two classes of sampling methods for solving
inverse problems within the Bayesian framework, namely Randomize-Then-Optimize (RTO) \cite{bardsley2014randomize}
and Langevin sampling methods \cite{roberts1996exponential,durmus2017nonasymptotic,durmus2018efficient}.
On one hand, RTO is deeply rooted within the variational framework as it solves
a perturbed optimization problem in order to generate one sample. On the other hand, Langevin methods
stem from the discretization of a stochastic differential equation (SDE) whose blue stationary distribution is the
distribution of interest.
The two classes of methods are similar in many ways, but they are
based on different principles and assumptions, and hence they lead to different densities. This
is illustrated in Figure \ref{fig:posterior_comparison}, which shows different solution
densities obtained from an observation $y = x+n$ where $n$ is standard normal and $x \in [a,b]$
is an unknown parameter. Adopting a uniform prior on $[a,b]$ leads to a truncated Gaussian
posterior. The Langevin approach requires smoothness and leads to a smooth approximate truncated
Gaussian posterior. In contrast, the RTO approach yields a density that is a mixture of a
truncated Gaussian distribution and a distribution on the boundary of the interval. The example
highlights some key differences: unlike the smooth approximate truncated Gaussian posterior,
the RTO density assigns non-zero probability to the boundary of the interval and has compact
support. This RTO density is associated with an implicit prior that is supported on $[a,b]$
and depends on the observation $y$. Consequently, it violates the typical Bayesian assumption 
that the prior is independent of the observation. Using somewhat unconventional terminology, we will refer to the RTO density as the RTO posterior.

As the example illustrates, the choice of prior has an important impact on the posterior $\pi_{\mathbbm{x}|\mathbbm{x}=y}(x)$.
A good prior is typically related to the nature of the inverse problem. In the Bayesian imaging
literature, there are many examples of priors that promote sparsity or piecewise regularity in
some transformed domain (e.g., involving the $l_1$ norm or the total-variation (TV) pseudo-norm
\cite{Rudin1992,Chambolle04,Louchet2013,Pereyra2016}). Priors can also come in the form of a
Markov random field \cite{MRF-MIT-2011}, a learned prior like a patch-based Gaussian, or a 
Gaussian mixture model \cite{Zoran2011,yu2011solving, Aguerrebere2014b,Teodoro2018scene,houdard2018high}. 
Choices pertaining to the prior are often informed by the resulting tractability of the posterior
\cite{Pereyra2016, durmus2018efficient,repetti_pereyra_2019,girolami2011riemann,chen2014stochastic}
or the ability to derive convergence guarantees \cite{Pereyra2016, durmus2018efficient,repetti_pereyra_2019,girolami2011riemann,chen2014stochastic}.
Moreover, recent progress on neural networks has spurred an interest in data-driven approaches
where a prior is learned from a large dataset $\{x_i\}_{i=1}^N \sim \pi_{\mathbbm{x}}(x)$ 
\cite{hurault2022proximal,laumont2022bayesian,holden2022bayesian,cai2023nfula,ho2020denoising}. 

In the following sections, we compare two sampling techniques:
Randomize-Then-Optimize (RTO) and the Moreau–Yoshida Unadjusted Langevin Algorithm (MYULA) \ \cite{durmus2018efficient}.
The latter finds frequent application in the field of imaging science. We begin by contrasting
the theoretical and practical aspects of these methods in Section~\ref{sec:theory_comp}, and we
quantitatively compare and evaluate RTO and MYULA in Section~\ref{sec:exp} on two classical
imaging inverse problems, namely deblurring and inpainting.
Finally, we conclude in Section~\ref{sec:conclusion} by summarizing the main findings and
discussing future research directions.


\section{Theory and Methods}\label{sec:theory_comp}

The posterior density $\pi_{\mathbbm{x}|\mathbbm{y}=y}(x)$ is often intractable and cannot be
computed in closed form. In such cases, the posterior density is explored using sampling
methods. In this section, we will study two such methods, namely Randomize-Then-Optimize (RTO)
and Langevin methods with a focus on MYULA. We will compare the two
classes of methods from a theoretical point of view, and discuss their
practical implications. We note that besides RTO and Langevin methods, there are many other
efficient sampling methods for high-dimensional densities based on Gibbs samplers, see e.g.\
\cite{vono2019split,pereyra2023split}. These are mainly designed to sample from high-dimensional
Gaussian distributions, and their adaptations to more complex target densities frequently involve
Langevin methods \cite{vono2019split}.

\subsection{Randomize-Then-Optimize (RTO) sampling method}

The RTO sampling method for Bayesian inverse problems was proposed in \cite{bardsley2014randomize}
as a way to approximate samples from a posterior distribution arising from a nonlinear inverse
problem with a Gaussian likelihood and a Gaussian prior. The basic idea is to perturb the
observation vector $y$ by noise and solve a MAP-like optimization problem to obtain a sample.
Specifically, if we consider the inverse problem in \eqref{eq:inv_pb} with a Gaussian likelihood
$\mathcal{N}(\mathcal{A}(x), \sigma^2 I)$ and a Gaussian prior $\mathcal{N}(x_0, S)$, the RTO
method generates a sample from the posterior distribution by solving the perturbed optimization
problem 
\begin{equation}\label{eq:optim_pb}
	x^\dag \in \argmin_{x}\, \left\{ \frac{1}{2\sigma^2} \|Q^T[\mathcal{A}(x) - \hat{y}] \|_2^2 + g(x) \right\}, \qquad g(x)=(x-x_0)^T S^{-1}(x-x_0),
\end{equation}
where the perturbed data $\hat{y}$ is drawn from the Gaussian pdf $\mathcal{N}(y, \sigma^2 I)$
and $Q$ is a matrix determining the tangent space of the posterior
mode. It results from a $QR$ decomposition of the Jacobian of $\mathcal{A}$, $J_{\mathcal{A}}$,
evaluated at $x^*=\argmin_u \|Au -y \|_2^2/2\sigma^2 + g(u)$. The idea of RTO in
the non-linear case is to generate samples from a Gaussian distribution and
to project them onto a manifold in a direction orthogonal to the tangent space of
$J_{\mathcal{A}}(x^*)$.
This process generates samples from an approximate posterior with
computable density $\hat{\pi}(x|y)$. Then samples from the exact posterior can be obtained
by adding a Metropolis-Hastings
step \cite{bardsley2014randomize,bardsley2018computational}. For a linear inverse problem, where $\mathcal{A}(x)=Ax$ and $Q=I$,
this procedure yields independent samples from the posterior and
reduces to a technique for Gaussian sampling proposed in \cite{PaY:10}.

Several extensions of RTO have been proposed in the literature. For example, in
\cite{wang2017bayesian}, RTO was extended to Laplace priors by
converting it to a standard Gaussian prior using a variable transformation.
Moreover, the RTO method has been extended to linear inverse problems
with implicit priors such as nonnegativity constraints \cite{BaF:12,bardsley2020mcmc}, polyhedral
constraints \cite{everink2023bayesian}, and more generally, implicit log-priors with a polyhedral
hypograph \cite{Everink_2023}. In the latter case, the function $g$ in \eqref{eq:optim_pb} is a
convex piecewise linear function that can be expressed as
\begin{equation}\label{eq:theory_prior}
 g(x) = \gamma \left( i_\setC(x) + \max_{i \in \setI} (c_i^T x + d_i) \right),
\end{equation}
where $\setC \subseteq \mathbb{R}^d$ is a polyhedral set, $\gamma>0$ is a constant, $\setI$ is a
finite index set, and $(c_i, d_i) \in \mathbb{R}^d \times \mathbb{R}$ for $i \in \setI$. This
class of implicit priors includes nonnegativity, Besov priors, and $l_1$-norm based priors
such as the anisotropic total variation (TV). As shown in \cite{everink2023bayesian}, RTO
generates samples from a well-defined probability density when $g$ is of the form
\eqref{eq:theory_prior} and $\rank(A) = d$. The resulting density assigns a positive probability
to low-dimensional sets that correspond to the faces of the polyhedral epigraph of $g$. In the
end, RTO reads
\begin{subequations}\label{eq:rto-split}
\begin{align}
	\mathbbm{z} &\sim \mathcal{N}(0, \sigma^2 I ) \label{eq:rtoa} \\
	\mathbbm{x} &= \operatorname{prox}_g^{\Sigma^{-1}} (A^\dagger(y + \mathbbm{z})) \ , \label{eq:rtob}
\end{align}
\end{subequations}

where $\Sigma^{-1} = A^T A/\sigma^2$, $A^\dagger = (A^T A)^{-1}A^T$ and $\operatorname{prox}_g^{\Sigma^{-1}}$
is the proximal operator with respect to the norm
induced by $\Sigma^{-1}$ defined for all $x \in\mathbb{R}^d$ by $\operatorname{prox}_g^{\Sigma^{-1}} (x) = \operatorname{argmin}_u \{g(u) + \frac{1}{2} \|x-u\|_{\Sigma^{-1}}^2\}$. In the case where $\rank(A) < d$, 
additional regularization may be necessary to guarantee the positive definiteness of $A^TA$ and the
existence of a well-defined posterior density. For example, this can be achieved by
adding a quadratic term of the form $\frac{\alpha}{2} \| x\|_2^2$ to the optimization problem in
\eqref{eq:optim_pb} \cite{everink2023bayesian}.

Although RTO is conceptually simple and can be used to sample from a wide range of densities,
we recall that it does not define a posterior distribution in a rigorous manner as its associated
implicit prior is observation-dependent and the Bayes' rule does not hold. RTO is rather anchored
in the sensitivity analysis framework, as it aims at quantifying the uncertainty resulting from
perturbations occurring in the data-space. Eventually, its underlying distribution describes
possible solutions given an observation.
From a more practical point of view, we emphasize that the cost of solving an optimization problem for
each sample can be prohibitive for high-dimensional problems. However, in practice it is not
necessary to solve the optimization problem to high accuracy to obtain useful samples.

\subsection{Langevin methods}

Langevin sampling methods arise from the Langevin stochastic differential equation (SDE), which reads 
\begin{equation}\label{eq:langevin}
	\textrm{d} x_t = \nabla \log \pi_{\mathbbm{x}|\mathbbm{y}=y}(x_t)\ \mathrm{d}t + \sqrt{2}\ \textrm{d} b_t \ ,
\end{equation}
where $(b_t)_{t \geq 0}$ denotes a $d$-dimensional Brownian motion, and the posterior density $\pi_{\mathbbm{x}|\mathbbm{y}=y}(x)$ is the target density. Under mild assumptions on $\pi_{\mathbbm{x}|\mathbbm{y}=y}(x)$, for any initial condition $x_0$,
\eqref{eq:langevin} admits a unique strong solution $(x_t)_{t \geq 0}$ with the target density as the unique stationary density \cite{roberts1996exponential}.
However, \eqref{eq:langevin} in general cannot be solved analytically, and we have to discretize it in order to solve it numerically. 

The unadjusted Langevin algorithm (ULA) is obtained by discretizing the Langevin SDE \eqref{eq:langevin} using the Euler--Maruyama scheme. This yields the homogeneous Markov chain
\begin{equation}\label{eq:ula}
	\mathbbm{x}_{k+1} = x_k + \delta \nabla \log \pi_{\mathbbm{x}|\mathbbm{y}=y}(x_k) + \sqrt{2\delta}\ \mathbbm{z}_{k+1},
\end{equation}
where $(\mathbbm{z}_{k+1})_{k \in \mathbb{N}}$ is a sequence of independent normal random variables, and $\delta >0$ is the discretization step-size that controls the
trade-off between accuracy and convergence-speed. Under mild assumptions on the target density,
this homogeneous Markov chain admits a unique stationary density $\pi_{\mathbbm{x}|\mathbbm{y}=y}^{\delta}(x)$
\cite{roberts1996exponential}, for which non-asymptotic bounds on the distance between the target
density $\pi_{\mathbbm{x}|\mathbbm{y}=y}(x)$ and the stationary density
$\pi_{\mathbbm{x}|\mathbbm{y}=y}^{\delta}(x)$ is provided in \cite{durmus2017nonasymptotic}. By
combining ULA with a Metropolis--Hasting step, one obtains the Metropolis-adjusted
Langevin algorithm (MALA) \cite{roberts1996exponential}, which produces samples from the target
posterior. ULA requires the target log-density to be
differentiable and is a popular method as it is straightforward to implement and scales
well with the dimension $d$ \cite{durmus2017nonasymptotic}. It has been extended to
nondifferentiable log-concave target densities by considering a surrogate density as proposed
in \cite{Pereyra2016,durmus2018efficient}. Specifically, if the target density can be expressed
as $\pi_{\mathbbm{x}|\mathbbm{y}=y}(x) \propto \exp (- f(x,y) - g(x))$ with nondifferentiable
$g(x)$ and with $f(\cdot,y)$ and $g$ both convex, the Moreau--Yoshida ULA (MYULA) is obtained
by using the Moreau--Yoshida envelope ${g}_{\alpha}$ of $g$ \cite{bauschke2017correction}
to construct a surrogate target density $\pi_{\mathbbm{x}|\mathbbm{y}=y}^{\alpha}(x) \propto
\pi_{\mathbbm{y}|\mathbbm{x}=x}(y)\  \pi_{\mathbbm{x}}^\alpha(x)$ with $\pi_{\mathbbm{x}}^\alpha(x) \propto
\exp(-g_\alpha(x))$. The function $g_\alpha$
is continuously differentiable with $1/\alpha$-Lipschitz gradient and such that $\nabla g_\alpha(x) = (x -\operatorname{prox}_g^\alpha (x))/\alpha$, where
$\operatorname{prox}_g^\alpha = \operatorname{prox}_g^{\alpha^{-1}I}$. The
resulting homogeneous Markov chain reads
\begin{subequations}
\begin{align}\label{eq:myula}
	\mathbbm{x}_{k+1} &= x_k + \delta \nabla \log \pi_{\mathbbm{x}|\mathbbm{y}=y}^{\alpha}(x_k)  + \sqrt{2\delta}\  \mathbbm{z}_{k+1} \\
	&= x_k - \delta [\nabla f(x_k, y) + \nabla g_\alpha (x_k)]  + \sqrt{2\delta}\ \mathbbm{z}_{k+1} \\
	&= \left(1-\frac{\delta}{\alpha}\right) x_k - \delta \nabla f(x_k, y) + \frac{\delta}{\alpha} \operatorname{prox}_g^\alpha (x_k) + \sqrt{2\delta}\ \mathbbm{z}_{k+1},
\end{align}
\end{subequations}
where $\operatorname{prox}_g^\alpha$ denotes the proximal operator associated with $g$ and parameter $\alpha>0$ \cite{bauschke2017correction}.
In other words, the transition Markov kernel is Gaussian and reads 
\begin{subequations}\label{eq:myula-split}
	\begin{align}
	\tilde{x}_{k+1}(x_k) &= \left(1-\frac{\delta}{\alpha}\right) x_k - \delta \nabla f(x_k, y) + \frac{\delta}{\alpha} \operatorname{prox}_g^\alpha (x_k), \label{ulaa} \\
	(\mathbbm{x}_{k+1}|\mathbbm{x}_{k} = x_k ) &\sim \mathcal{N}(\tilde{x}_{k+1}(x_k), 2\delta I).\label{ulab}
\end{align}
\end{subequations}
Non-asymptotic bounds on the distance between the target density $\pi_{\mathbbm{x}|\mathbbm{y}=y}(x)$ and the
stationary density associated with MYULA $\pi_{\mathbbm{x}|\mathbbm{y}=y}^{\alpha, \delta}(x)$
are provided in \cite{durmus2018efficient}.

MYULA is conceptually simple. The evaluation of the proximal operator
$\operatorname{prox}_g^\alpha$ requires the solution of a convex optimization problem, and
hence the cost of a sample can be high if no closed-form expression is available. However, as
shown empirically in \cite{durmus2018efficient}, evaluating the proximal operator inaccurately
can still produce good results.

\subsection{Conceptual differences between RTO and MYULA}

We now compare RTO and Langevin methods from a theoretical point of view while paying special attention to models with non-differentiable priors. Specifically, we will focus on the RTO method proposed in \cite{Everink_2023,everink2023bayesian} and MYULA \cite{Pereyra2016,durmus2018efficient}. \Cref{tab:th_comp} lists the main conceptual differences, which we will now explain in more details.
To simplify this comparison, we focus on a Gaussian likelihood as the behavior of RTO can be characterized in this setting.

\begin{table}[t]
	\centering
	\begin{tabular}{|l|l|l|}
		\hline
		& RTO & MYULA
		\\
		\hline 
		\hline
		Log-prior   & polyhedral hypograph & concave
		\\
		\hline
		Target density & implicit form  & explicit form
		\\
		\hline 
		Random perturbation & in data space & in image space 
		\\
		\hline 
		Samples & independent & correlated 
		\\
		\hline 
		Burn-in & no & required 
		\\
		\hline
		Sample generation cost & high: computing \eqref{eq:rtob} & low: computing \eqref{ulaa}
		\\
		\hline
		Parameter selection & online & offline 
		\\
		& hierarchical approach & empirical approach 
		\\
		\hline
		Parallelization & yes & possible but difficult 
		\\
		\hline
	\end{tabular}
	\caption{Comparisons between RTO and MYULA from a theoretical point of view for a Gaussian likelihood.}
	\label{tab:th_comp}
\end{table}

Firstly, it is important to note that the two methods are based on different assumptions. RTO's analysis relies on
the assumption that the likelihood is Gaussian and that the log-prior has a polyhedral hypograph. In contrast,
MYULA is compatible with more general posteriors that are formed from a continuously differentiable log-concave
likelihood with a Lipschitz gradient and a log-concave prior. Because of the discretization of the Langevin SDE
defined in \eqref{eq:langevin} and the use of a surrogate log-prior density, MYULA does not actually produce samples
from the target posterior but from a surrogate posterior density. The distance between the surrogate density and
the target density can be quantified in a non-asymptotic sense. However, the samples generated by RTO are from a
pdf in an implicit form, which is also different from the target posterior. 

Both RTO and MYULA involve a stochastic perturbation to produce a sample, but at different stages.
RTO applies a perturbation in the data space before solving an instance of the minimization problem, as shown in \eqref{eq:rto-split}. 
The operation in \eqref{eq:rtob} in fact corresponds to the generalization of an oblique projection and tends to
concentrate the probability mass in some areas of the space. It partially explains why RTO can assign a strictly positive
probability mass to low-dimensional subsets and why RTO samples can be sparse. In MYULA, the perturbation is
applied in the image space after performing an explicit gradient descent step with respect to the surrogate
posterior, see \eqref{eq:myula-split}. That is why Langevin-based methods can be interpreted as a perturbed gradient descent
\cite{welling2011bayesian}. In MYULA, due to the use of the surrogate model to overcome non-differentiable prior
we also need perform a proximal operation but only on the prior, see \eqref{ulaa}. The computation of this proximal operation is often much cheaper than the one in RTO, since the forward operator $A$ is not involved. The accuracy of computing $\operatorname{prox}_g^\alpha$ has an impact on the quality of samples, which
is investigated in \Cref{sec:deblurring}. Because the density sampled by MYULA is continuous on the whole image space,
MYULA is then unable to enforce constraints.

Comparing the computational costs, we need to solve an optimization problem, i.e.,
computing \eqref{eq:rtob}, in order to generate an RTO sample.
However, with MYULA we obtain one sample at each iteration that involves the
computation of a proximal operator, i.e., computing \eqref{ulaa}. In RTO, as the perturbations are
independent, samples are independent. Furthermore, we can parallelize RTO to
generate several samples at the same time. MYULA, as all Markov chain Monte Carlo
(MCMC) methods, require a burn-in period in order for the Markov chain to enter
its stationary regime and actually
produce samples from its stationary density. The length of the burn-in period is
unknown a priori. In addition,
samples generated from MYULA as well as other Langevin methods are correlated
as they are generated by using the previous
sample. Although expected, it can have a negative impact on the
accuracy of the Monte Carlo estimates. 
Furthermore, to address the common convergence speed issues of MCMC
methods and accelerate their convergence, Jacob et al. proposed in \cite{jacob2020unbiased}
to compute Monte Carlo estimates by coupling
Markov chains running in parallel.
However, their practical implementations in a very high-dimensional setting
remains a challenge.

Finally, we will compare the flexibility on the extension to model parameter
selection. Often the posteriors include some unknown parameters, e.g. $\sigma$ in
\eqref{eq:optim_pb}, $\gamma$ in \eqref{eq:theory_prior} and $\alpha$ in
\eqref{eq:myula}. The choice of those parameters can have a great impact on the
posterior \cite{molina1999bayesian,pereyra2015maximum}. RTO allows us to perform
automatic parameter selection by considering a hierarchical model, i.e. we add
prior distributions on the hyperparameters $\lambda=1/\sigma^2$ and $\gamma$ and
consider the posterior
\begin{equation}
\pi_{\mathbbm{x},\bblambda,\bbgamma|\mathbbm{y}=y}(x,\lambda,\gamma)=\pi_{\mathbbm{x}|\mathbbm{y}=y,\bblambda=\lambda,\bbgamma=\gamma}(x)\pi_{\bblambda}(\lambda)\pi_{\bbgamma}(\gamma).
\end{equation}

To utilize the conjugacy with Gaussian distributions, we impose $\Gamma$-distributions
for both $\bblambda$ and $\bbgamma$. The negative-logarithm of the conditional
distribution $\pi_{\mathbbm{x}|\mathbbm{y}=y,\bblambda=\lambda,\bbgamma=\gamma}$
is proportional to
\begin{equation}
\lambda \|\mathcal{A}(x) - \hat{y} \|_2^2 + g(x)
\end{equation}
with $g(x)$ given in \eqref{eq:theory_prior}. 
\Cref{alg:augmented_rto} summarizes the procedure to sample from such a hierarchical model within the RTO
framework.

\begin{algorithm}
	\caption{RTO Hierarchical Gibbs sampler for $(\mathbbm{x},\bblambda,\bbgamma)$}\label{alg:augmented_rto}
	\begin{algorithmic}
	\Require $N \in \mathbb{N}^*, \ x_0 \in \mathbb{R}^d. $
	\Ensure{$(x_k)_{k \in \{1, ..., N\} }$}
	\For{$k=1 \mathrm{\ to \ } N$}
	\State{\underline{Step 1:} Sample $\lambda_k \sim \pi_{\bblambda | \mathbbm{x}=x_{k-1}, \mathbbm{y}=y}(\lambda)$ and $\gamma_k \sim \pi_{\bbgamma | \mathbbm{x}=x_{k-1}, \mathbbm{y}=y}(\gamma)$, and both follow $\Gamma$-distributions.}  
	\State{\underline{Step 2:} Sample $x_k \sim \pi_{\mathbbm{x}|\bblambda=\lambda_k,\bbgamma=\gamma_k,\mathbbm{y}=y}(x)$ by RTO.}
	\EndFor
	\end{algorithmic}
\end{algorithm}

In principle, with Langevin-based methods, the selection of posterior
parameters could also be performed by considering a hierarchical approach. However,
this approach is not feasible for dealing with high-dimensional problems, since
in each iteration MYULA would possibly need many iterations in order to enter its
stationary regime to obtain a sample from
$\pi_{\mathbbm{x}|\bblambda=\lambda_k,\bbgamma=\gamma_k,\mathbbm{y}=y}(x)$.
Recently, Vidal et al. proposed in \cite{vidal2020maximum,de2020maximum} to perform parameter
selection based on maximum likelihood estimation,
which can simultaneously estimate multiple model parameters and is theoretically well-founded.
However, the procedure has to be implemented offline, i.e., the parameters have to be
predetermined before sampling.

\section{Experimental study}\label{sec:exp}

In this section, we aim at numerically investigating the differences between RTO and MYULA methods for
sampling a high-dimensional posterior model. To do so, we tackle two kinds of imaging inverse problems 
described by \eqref{eq:inv_pb}: deblurring and inpainting. In both cases, the forward operator $A$ is known and
linear. 

\begin{figure}
    \centering
    \begin{tabular}{cc}
        \includegraphics[width=0.3\textwidth]{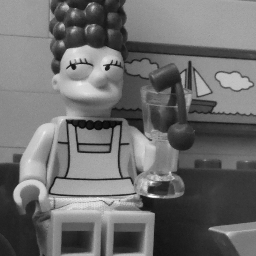} &
        \includegraphics[width=0.3\textwidth]{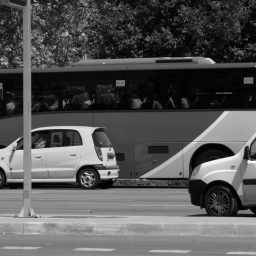}
        \\
        \texttt{Simpson} & \texttt{Traffic} 
    \end{tabular}
    \caption{Original images used for the deblurring and inpainting experiments.}
    \label{fig:original_images}
\end{figure}

\Cref{fig:original_images} shows both 256-by-256 test images with an intensity range $[0,1]$ that were used in this
section. These images have been selected because of their differences in content and level of details. For example,
\texttt{Traffic} presents a mix of piece-wise constant and more textured areas. We consider the prior of the form
$p(x) \propto \exp (-g(x))$ with 
\begin{equation}\label{eq:gnu}
g(x)= \gamma \|\nabla x \|_{1, 1}+i_\setC (x)
\end{equation}
and $\setC = [0, 1]^d$.
It is a non-differentiable log-concave prior which promotes images with sparse gradients.
We compare the results obtained by RTO with the results of MYULA.
Note that both RTO and MYULA sample their target posterior densities approximately, and their
corresponding approximations are different.

We generate 1000 RTO samples. To solve the minimization problem \eqref{eq:optim_pb} with $g(x)$ defined in
\eqref{eq:gnu} in RTO, we apply the alternating direction method of multipliers (ADMM) with the stopping criterion
suggested in \cite[Eq (3.12)]{boyd2011distributed}. According to the theories in \cite{Everink_2023,everink2023bayesian},
to guarantee a well-defined posterior density we need to add a term $\frac{\alpha}{2}\|x\|_2^2$ for the rank-deficient
$A$, which is the case here, and we set $\alpha=10^{-8}$ to ensure little impact on the results.

According to $g(x)$ defined in \eqref{eq:gnu}, MYULA targets the surrogate posterior density $\pi_{\alpha_1, \alpha_2} (x|y)$: 
\begin{equation}\label{eq:deblurring_tv_myula}
\pi_{\mathbbm{x}|\mathbbm{y}=y}^{\alpha_1 , \alpha_2}(x) \propto \exp \left(-\frac{1}{2\sigma^2} \|Ax -y \|_2^2 - [\gamma \|\nabla . \|_{1, 1}]_{\alpha_1} (x) - [i_c]_{\alpha_2} (x) \right) \ , 
\end{equation}
where $[\gamma \|\nabla . \|_{1, 1}]_{\alpha_1}$ and $[i_c]_{\alpha_2}$ denote the Moreau-Yoshida envelopes of the two terms in $g$, respectively. 
To apply MYULA defined in \eqref{eq:myula}, we need $\operatorname{prox}_{\gamma \|\nabla . \|_{1, 1}}^{\alpha_1}$.
However, since $\operatorname{prox}_{\gamma \|\nabla . \|_{1, 1}}^{\alpha_1}$ has no closed-form expression,
we apply the first-order primal-dual algorithm introduced in \cite{chambolle2011first} to estimate it. In the
following, we set the number of iterations to estimate this proximal operator to $n_{pd} = 50$. We initialize MYULA
at the data $y$ and run MYULA for $2.5\times 10^5$ iterations after $2.5\times 10^4$ burn-in iterations. As suggested in
\cite{durmus2018efficient}, we set $\alpha_1 = \alpha_2 = 1/(\|A^T A \|/\sigma^2)$ and $\delta = 1/(2/\alpha_1 + 2/\alpha_2 + 2\|A^T A \|/\sigma^2 )$.

To quantitatively evaluate the reconstruction quality, we use the peak signal-to-noise ratio (PSNR) and the
structural similarity index measure (SSIM) \cite{wang2004image,wang2009mean}.

\subsection{Deblurring}\label{sec:deblurring}

\begin{figure}
	\centering
	\begin{tabular}{cc}
		\includegraphics[width=0.3\textwidth]{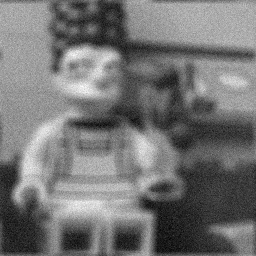} &
		\includegraphics[width=0.3\textwidth]{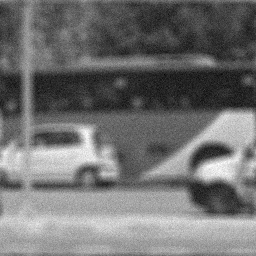} 
		\\
		PSNR=21.75/SSIM=0.47 & PSNR=19.90/SSIM=0.37 
	\end{tabular}
	\caption{\textit{(Deblurring)} degraded images.}
	\label{fig:deblurring_obs}
\end{figure}

In this section, images are degraded by a $9\times 9$ uniform blurring kernel and
additive white Gaussian noise with zero mean and variance $\sigma^2 = 0.001$. 
We assume the periodic boundary condition, which turns out that $A$ can be diagonalized by the Fourier transform.
In 
\Cref{fig:deblurring_obs} we show the degraded images.

For the stopping criteria in ADMM for RTO, we set the tolerance of the primal and dual residuals to
$\texttt{tol}=10^{-4}$ and the maximum iteration number to $\texttt{maxiter}= 2000$. In addition, we set the
regularization parameter $\gamma$ in $g(x)$ to $\gamma = 5$. For MYULA, we use $\gamma=10$.

\begin{remark}
As explained in \Cref{sec:theory_comp}, regularization parameters can have significant impacts on the results. For
the regularization parameter $\gamma$ in RTO, we use the parameter achieving the largest PSNR when solving the
unperturbed optimization problem. It generally delivers good results in the RTO framework. It means that the
maximum-a-posteriori (MAP) point estimate can be a good reference for selecting $\gamma$ and highlights the
connection between optimization and sampling within the RTO framework.
Finding a good $\gamma$ for MYULA is more complicated, since we found that the regularization parameter
generating a MAP with high PSNR and SSIM scores for the unperturbed optimization problem often leads to too
noisy samples in MYULA. 
Consequently, we had to run several Markov chains with different regularization parameters $\gamma$ in order to find a
good value, which is much more time-consuming compared to RTO.
\end{remark}

\begin{figure}
\centering
	\begin{tabular}{cccc}
		& RTO & MYULA & MAP \\
		\CenteredVcell{0.3\textwidth}{\texttt{Simpson}} &
		\includegraphics[width=0.3\textwidth]{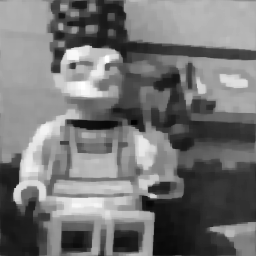} &
		\includegraphics[width=0.3\textwidth]{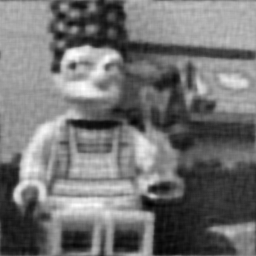} &
		\includegraphics[width=0.3\textwidth]{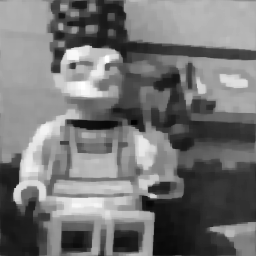} 
		\\
		& \textbf{PSNR=25.69/SSIM=0.79} & PSNR=24.98/SSIM=0.72 & PSNR=25.17/SSIM=0.77 
		\\
		\CenteredVcell{0.3\textwidth}{\texttt{Traffic}} &
		\includegraphics[width=0.3\textwidth]{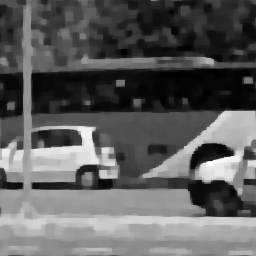} &
		\includegraphics[width=0.3\textwidth]{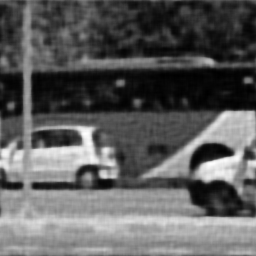} &
		\includegraphics[width=0.3\textwidth]{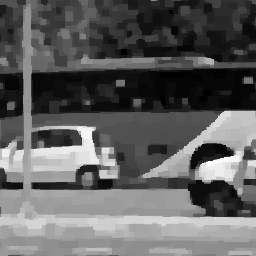} 
		\\
		& \textbf{PSNR=24.00/SSIM=0.66} & PSNR=23.18/SSIM=0.63 & PSNR=23.69/SSIM=0.65 
	\end{tabular}
	\caption{\textit{(Deblurring)} MMSE estimates computed respectively with RTO and MYULA together with MAP estimates.}
	\label{fig:deblurring_tv_mmse}
\end{figure}

\textbf{Results and discussions:} In \Cref{fig:deblurring_tv_mmse} we show the minimum mean square error (MMSE)
estimates of RTO and MYULA together with the MAP estimates for comparison. We can see that MMSE of RTO
achieves the highest PSNR and SSIM scores. Visually it resembles the MAP estimate with the stair-casing artifacts
due to the use of TV regularization.
However, we do not observe the stair-casing artifacts for the MMSE estimate computed by MYULA, but   
a grid pattern ruins the restoration.

\begin{figure}
\centering
	\begin{tabular}{ccc}
		\CenteredVcell{0.3\textwidth}{RTO}
		\includegraphics[width=0.3\textwidth]{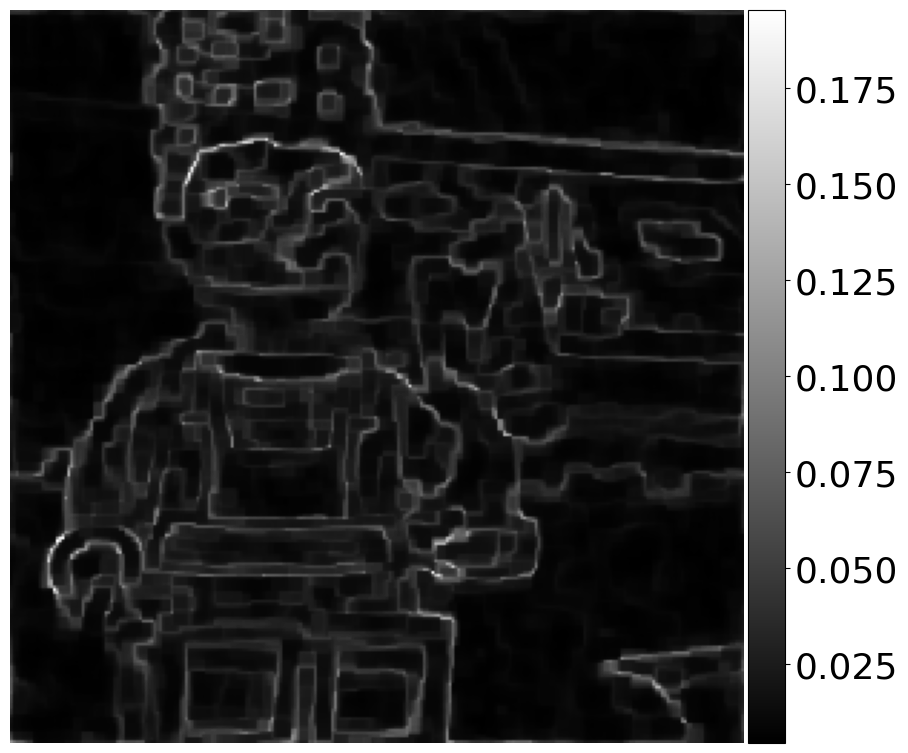} &
		\includegraphics[width=0.3\textwidth]{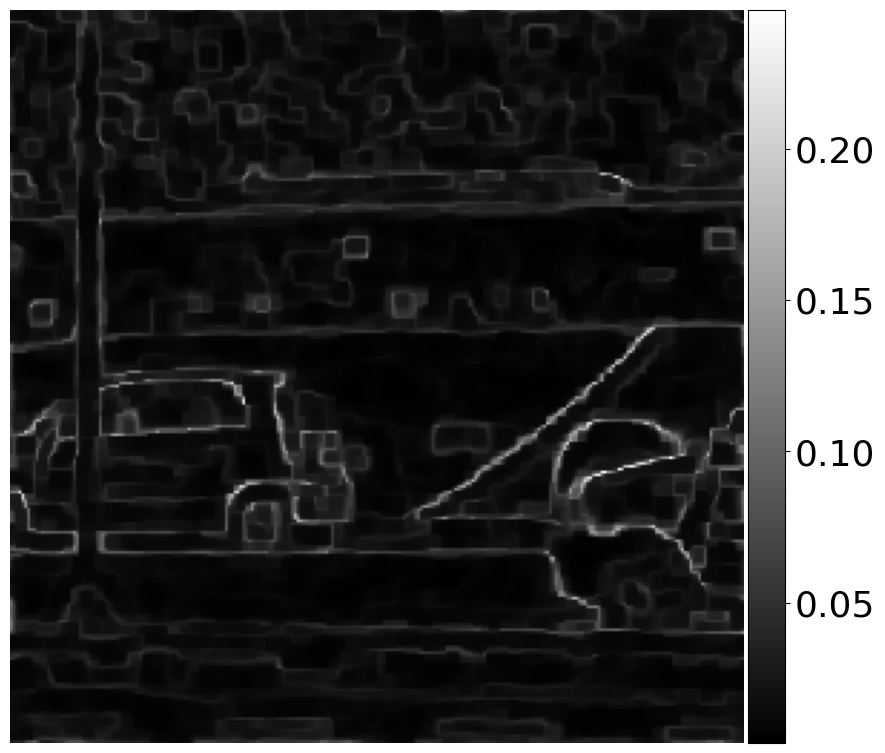}
		\\
		\CenteredVcell{0.3\textwidth}{MYULA}
		\includegraphics[width=0.3\textwidth]{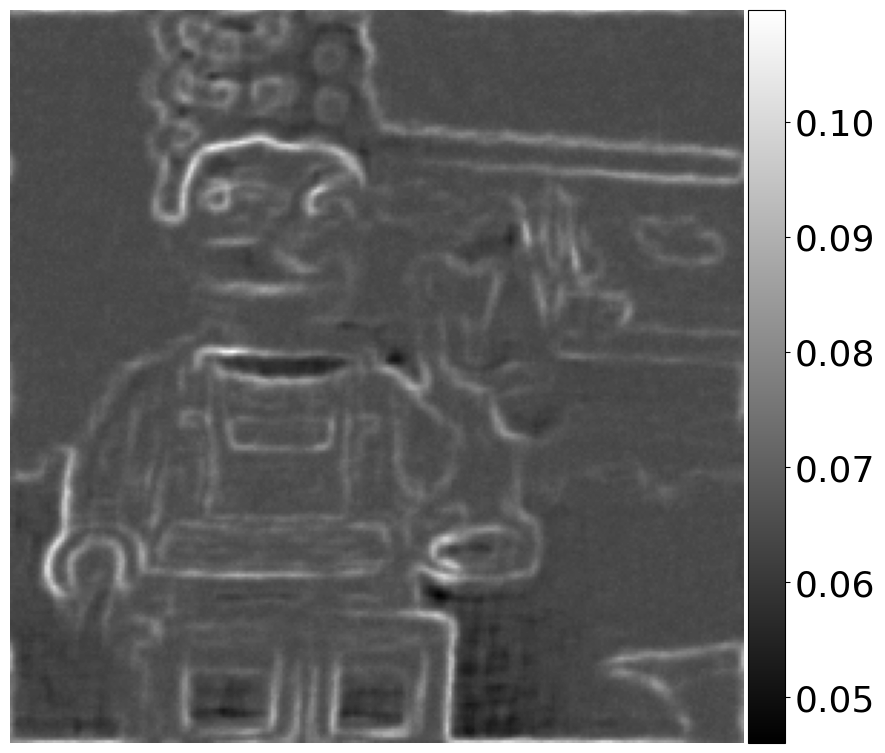} &
		\includegraphics[width=0.3\textwidth]{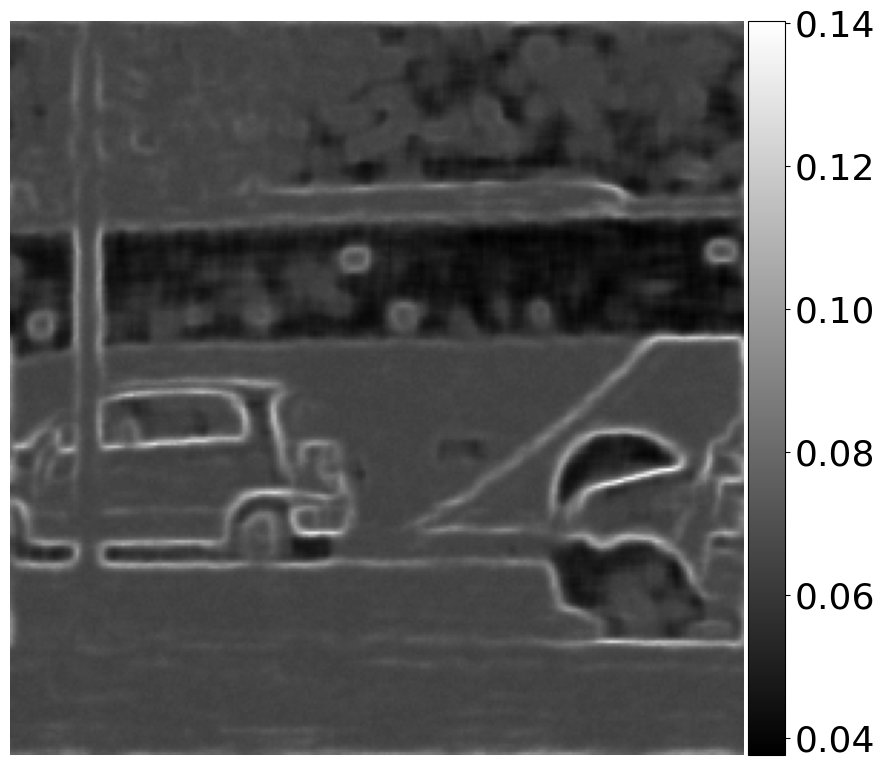}
	\end{tabular}
	\caption{\textit{(Deblurring)} Marginal posterior standard deviation computed with the samples generated by RTO and MYULA.}
	\label{fig:deblurring_tv_std}
\end{figure}

\Cref{fig:deblurring_tv_std} shows the standard deviation maps of RTO and MYULA, respectively. In the results of
both methods we observe higher uncertainties around edges, which is coherent as high frequency information is more
challenging to restore. However, the standard deviation of MYULA exhibits a more spread out uncertainty: edges are
less uncertain than the ones of RTO whereas the constant regions have larger standard deviations than those of RTO.
These differences relate to where the perturbation is performed.
In RTO we perturb the data before computing the proximal operator, whereas in MYULA the perturbation takes
place after taking a gradient step in the image space.

\begin{figure}
    \centering
	\begin{tabular}{ccc}
		& \texttt{Simpson} & \texttt{Traffic}
		\\
		\CenteredVcell{0.3\textwidth}{RTO} &
		\includegraphics[width=0.3\textwidth]{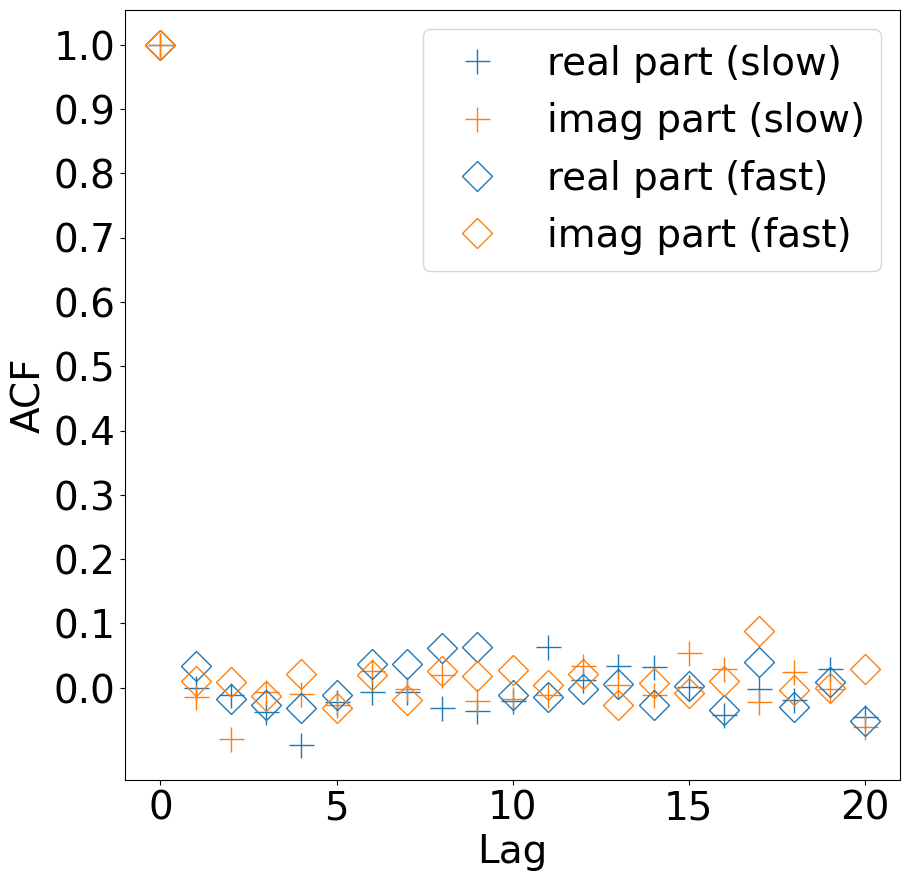} &
		\includegraphics[width=0.3\textwidth]{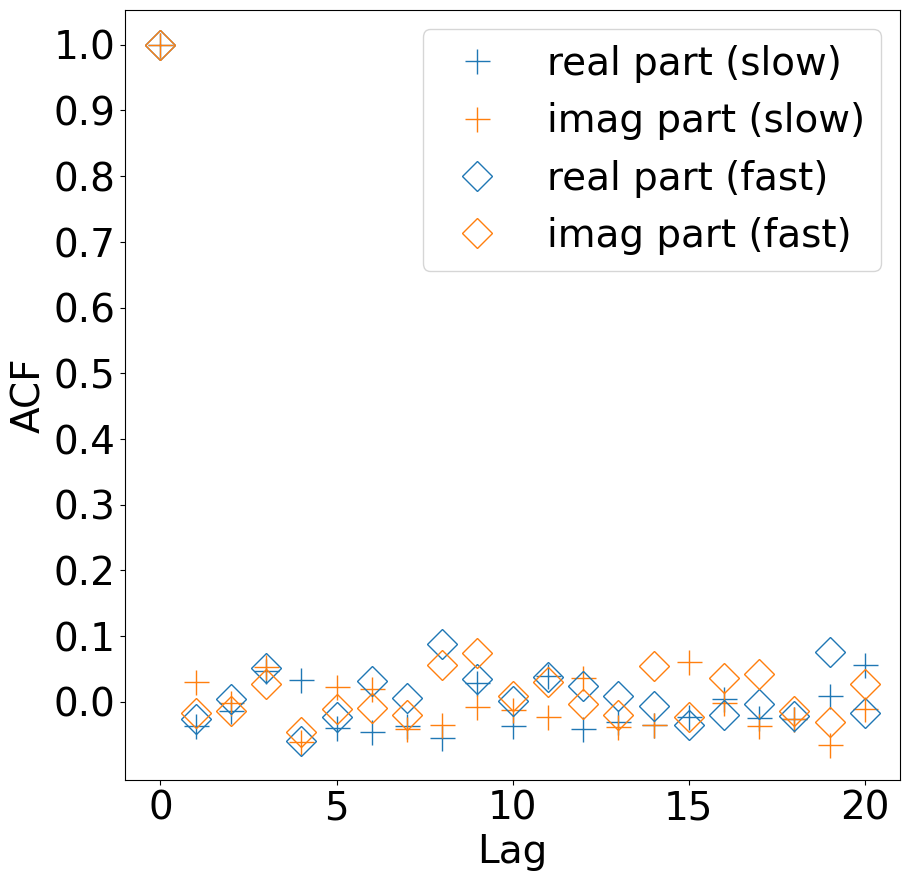}
		\\
		\CenteredVcell{0.3\textwidth}{MYULA} &
		\includegraphics[width=0.3\textwidth]{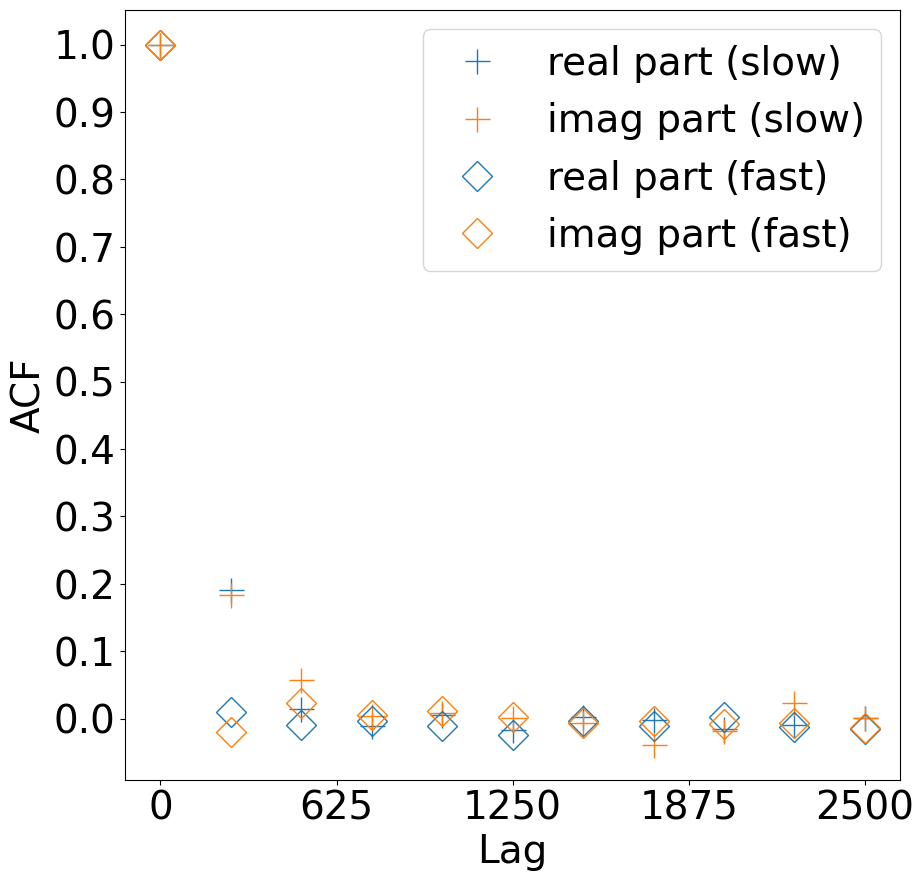} &
		\includegraphics[width=0.3\textwidth]{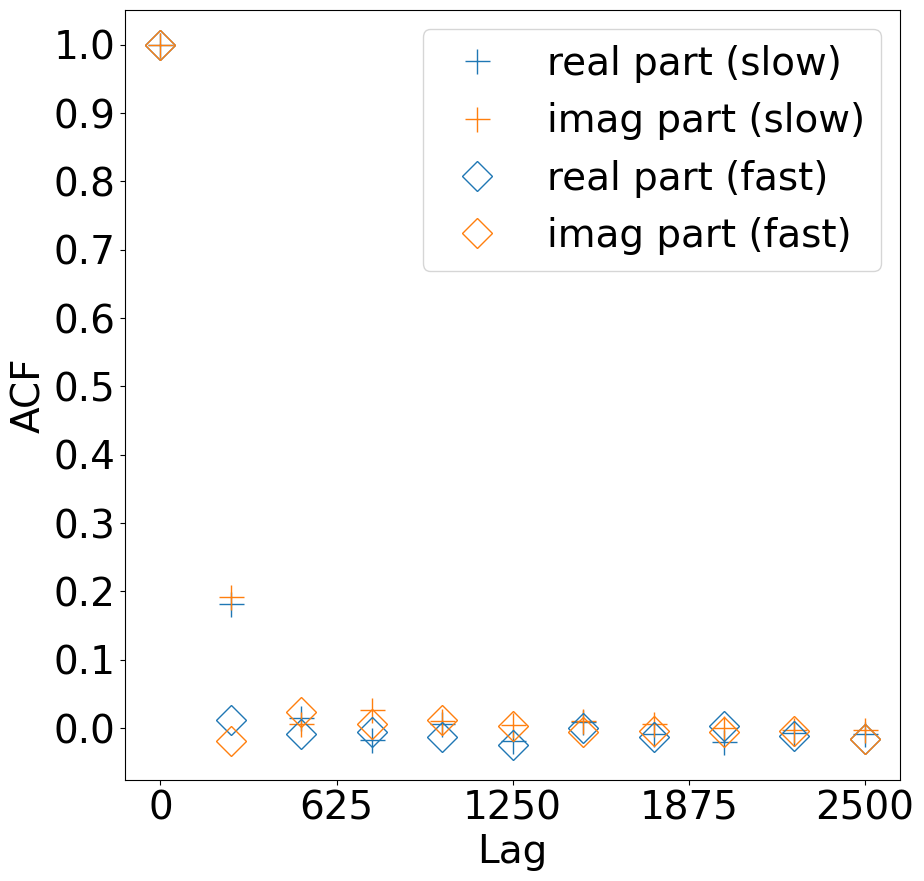}
	\end{tabular}
	\caption{\textit{(Deblurring)} The comparison of ACFs of RTO and MYULA on both test images.}
	\label{fig:deblurring_tv_acf}
\end{figure}

To check the correlation among the samples, we compute  
the sample auto-correlation functions (ACFs), which measure how fast samples become uncorrelated. A fast decay of ACFs indicates no linear dependency among the samples and
is also interpreted as a short mixing time for Markov chains. It comes with accurate Monte-Carlo estimates. As
images are high-dimensional objects, directly estimating the $d$-dimensional ACFs is not realistically doable.
However, the convergence speeds can be inferred from the posterior covariance matrix \cite{pereyra2020accelerating}.
We assume that the posterior covariance shape is mostly determined by the likelihood. Therefore, we approximate
the posterior covariance  using the directions provided by the diagonalization basis of the forward operator $A$,
i.e., the Fourier basis, for the deblurring experiments as in \cite{laumont2022bayesian}.
\Cref{fig:deblurring_tv_acf} shows the evolution of the estimated ACFs both of RTO and the complete Markov
chain of MYULA. ACFs of RTO samples immediately 
drop to zero, which shows that RTO samples are independent, while samples generated by MYULA only become
uncorrelated after $500$ iterations. It means 
that in order to get $1000$ uncorrelated samples,
we need to run MYULA for $5\times10^5$ iterations. Note that the number of iterations before MYULA samples get
uncorrelated actually depends on the inverse problem we deal with, and cannot evaluate a-priori but only after
running the Markov chain.
Furthermore, MYULA requires a burn-in phase that corresponds to the time required by the chain to enter in its
stationary regime, i.e., when samples generated by the Markov chain are sampled from $\pi_{\delta,\alpha_1, \alpha_2} (x|y)$. The number of iterations in burn-in also cannot assess a-priori but only after running the chain.

\begin{figure}
	\centering
	\begin{tabular}{cccc}
		\CenteredVcell{0.25\textwidth}{RTO}
		\includegraphics[width=0.25\textwidth]{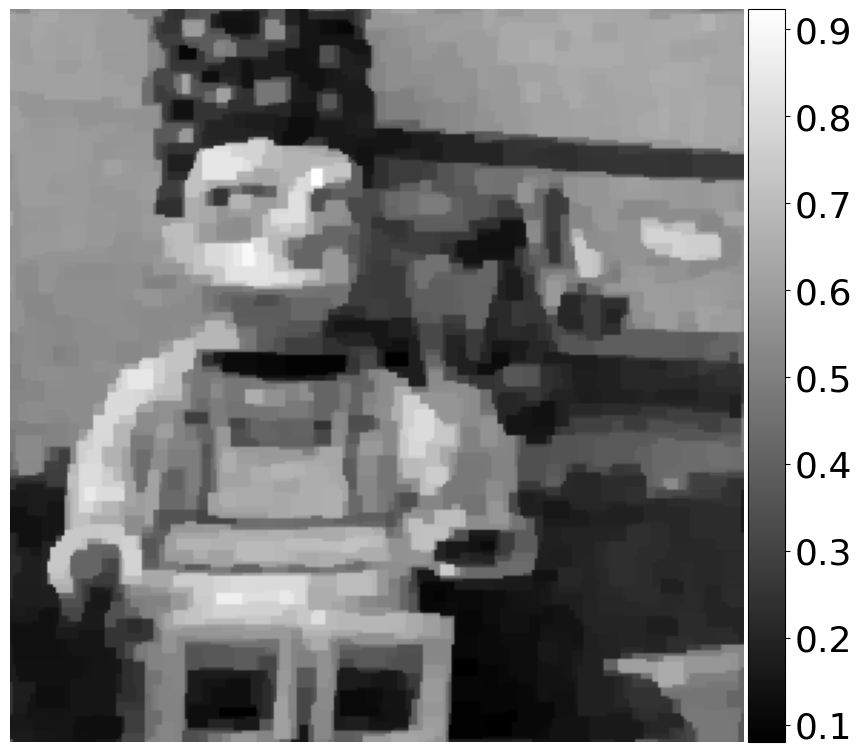}
		&
		\includegraphics[width=0.25\textwidth]{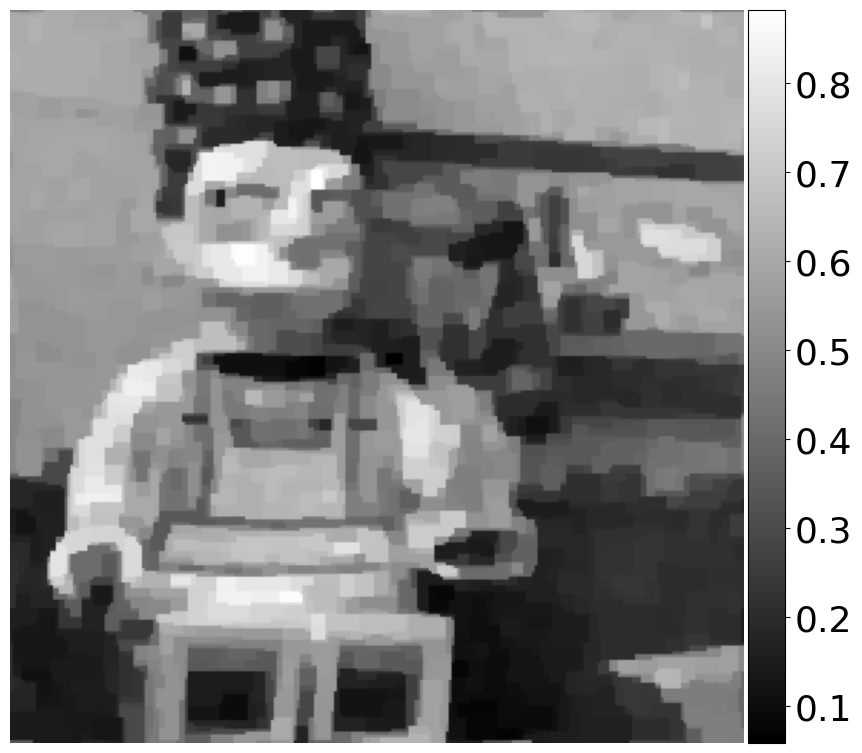}
		&
		\includegraphics[width=0.25\textwidth]{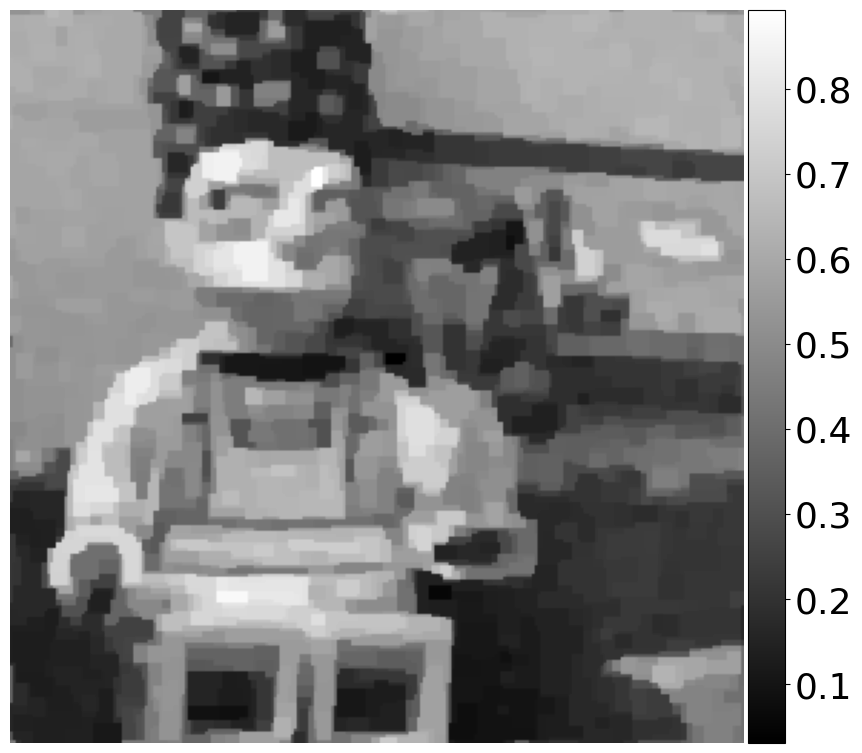}
		\\
		PSNR=24.76/SSIM=0.74 & PSNR=24.72/SSIM=0.74 & PSNR=24.72/SSIM=0.74
		\\
		\CenteredVcell{0.25\textwidth}{MYULA}
		\includegraphics[width=0.25\textwidth]{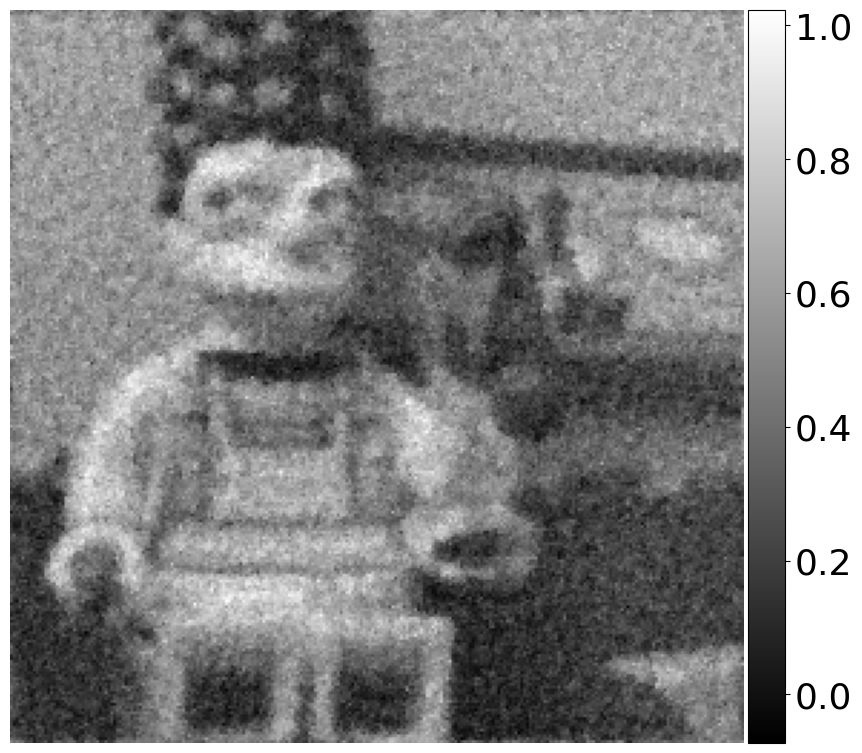}
		&
		\includegraphics[width=0.25\textwidth]{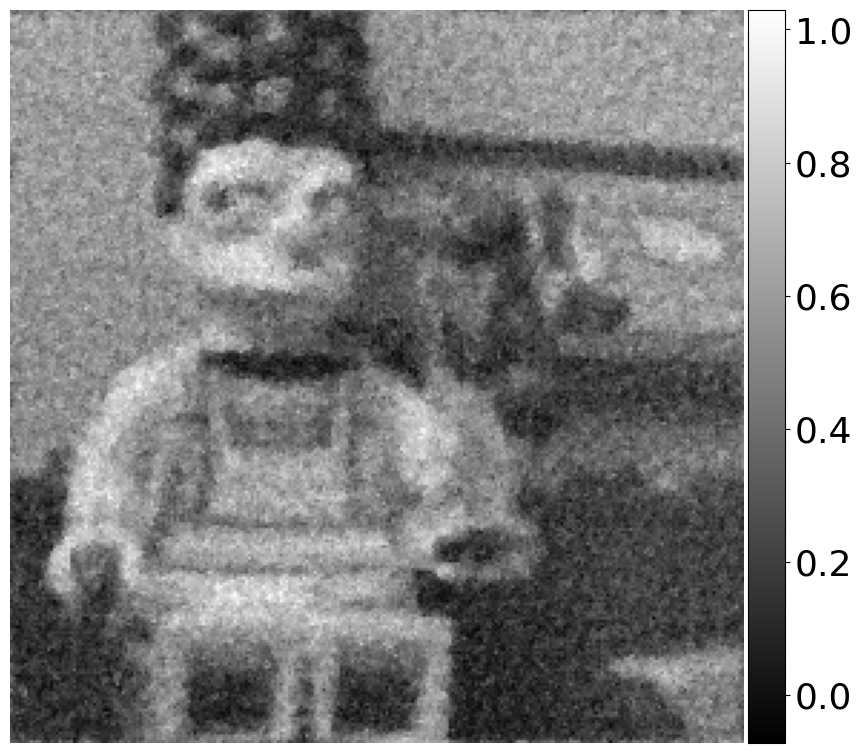}
		&
		\includegraphics[width=0.25\textwidth]{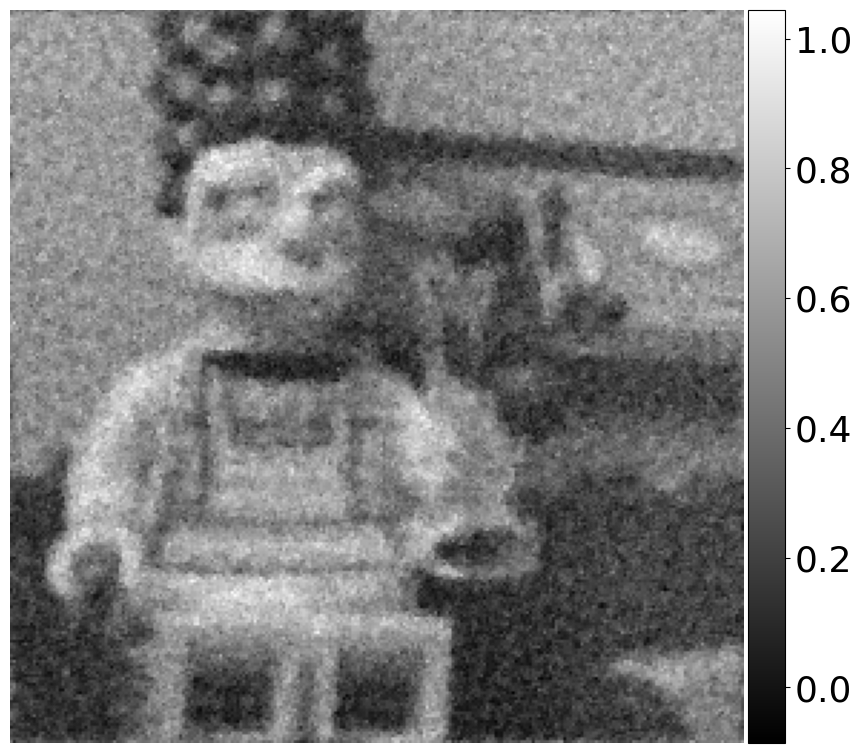}
		\\
		PSNR=21.17/SSIM=0.37 & PSNR=21.18/SSIM=0.37 & PSNR=21.18/SSIM=0.37
		
	\end{tabular}
	\caption{\textit{(Deblurring)} Samples generated by RTO and MYULA.}
	\label{fig:deblurring_tv_samples}
\end{figure}

In \Cref{fig:deblurring_tv_samples} we show a few samples generated by RTO and MYULA, respectively.
It is clear that MYULA samples are much more noisy than the ones of RTO. The reason is that RTO samples are in
fact MAP estimates with a perturbed observation. On the other hand, MYULA samples are generated by a gradient
descent step with added noise. In addition, we can see that MYULA samples do not belong to $\setC$, contrarily
to RTO samples.

\begin{figure}
	\centering
	\begin{tabular}{ccc}
		& \texttt{Simpson} & \texttt{Traffic}
		\\
		\CenteredVcell{0.3\textwidth}{RTO} &
		\includegraphics[width=0.3\textwidth]{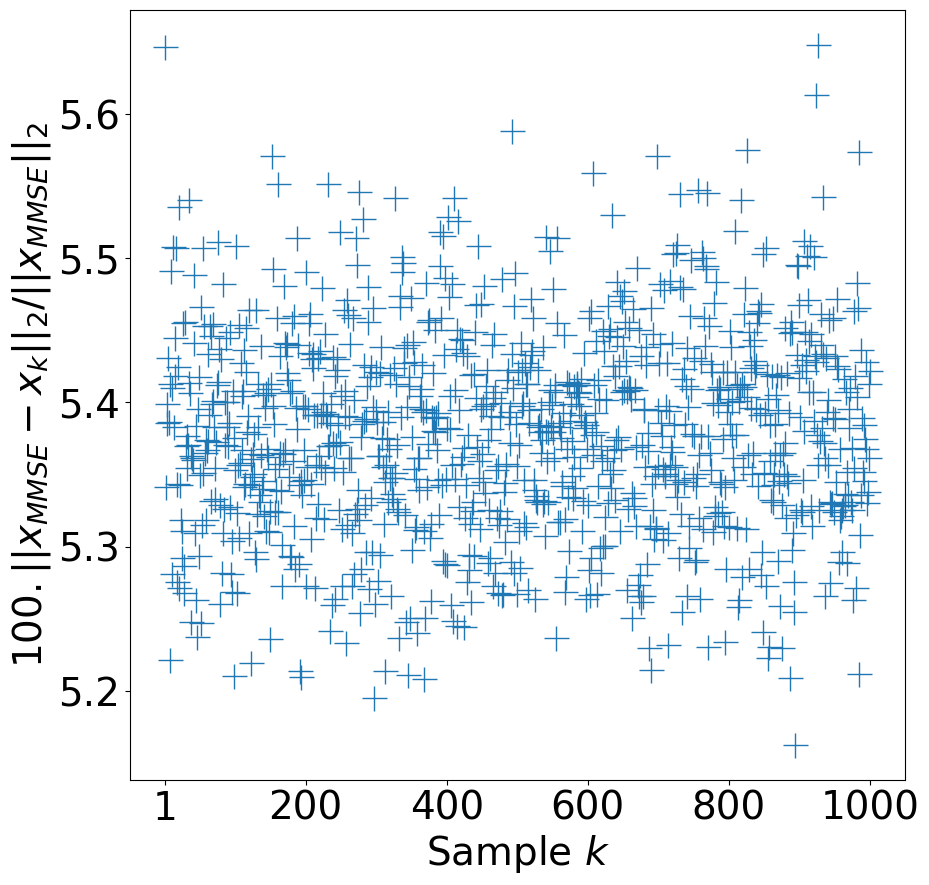} &
		\includegraphics[width=0.3\textwidth]{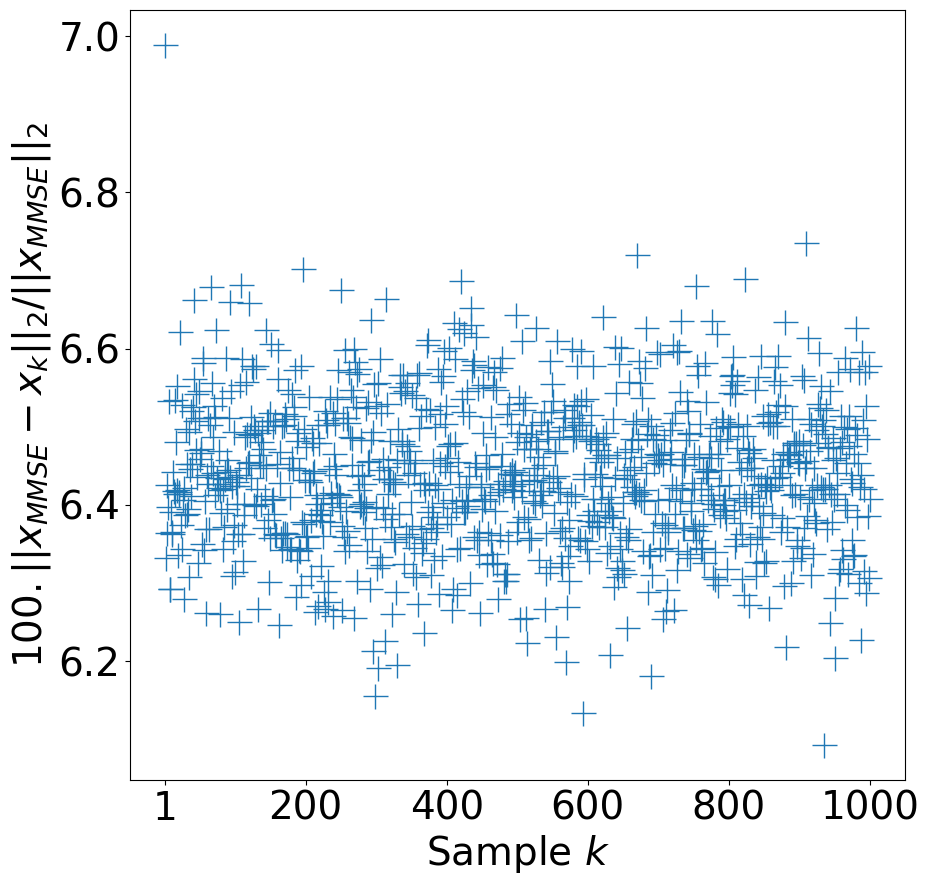}
		\\
		\CenteredVcell{0.3\textwidth}{MYULA} &
		\includegraphics[width=0.3\textwidth]{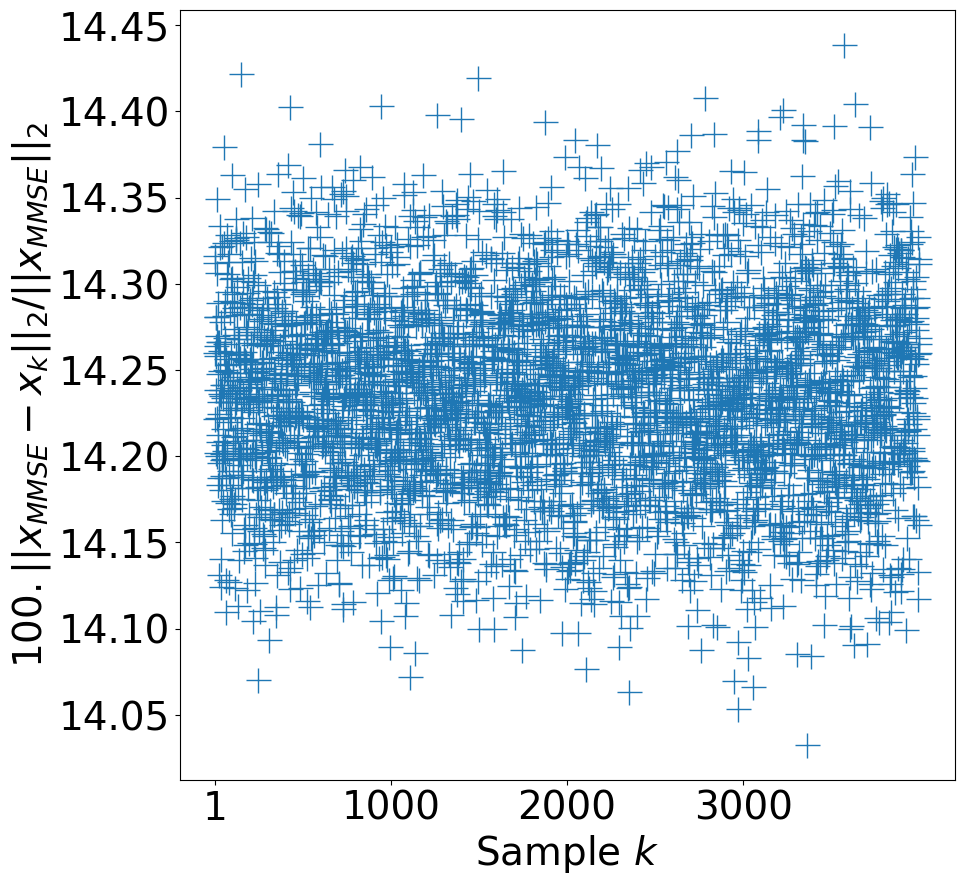} &
		\includegraphics[width=0.3\textwidth]{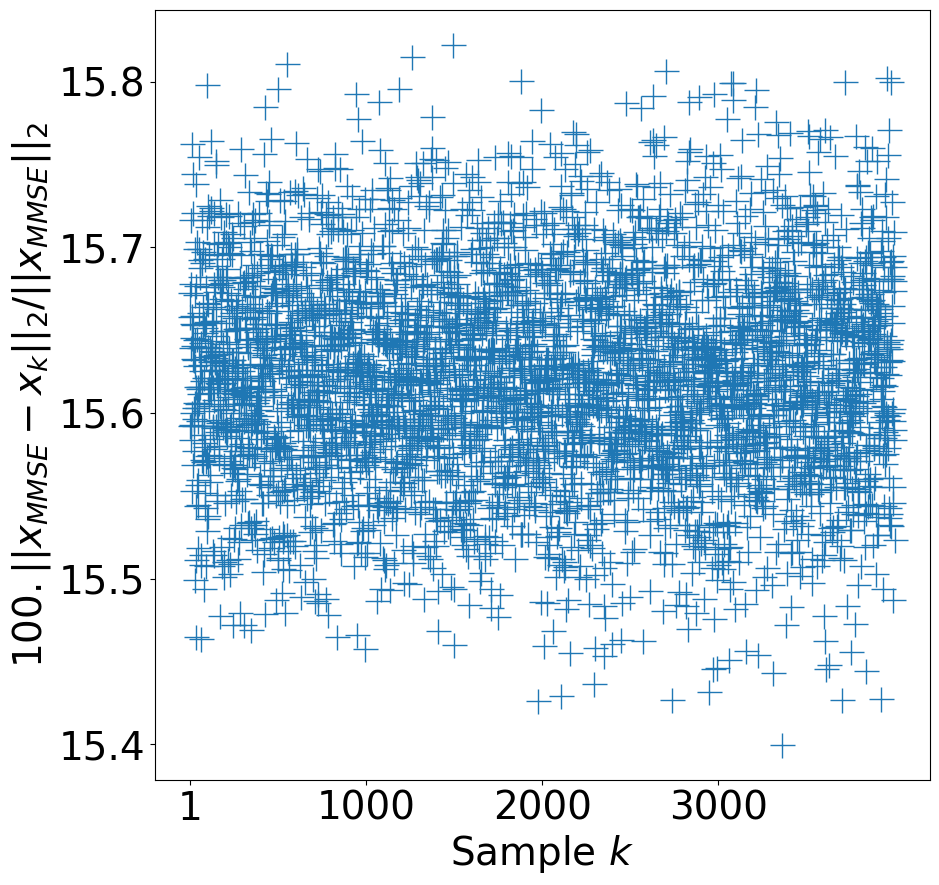}
	\end{tabular}
	\caption{\textit{(Deblurring)} Relative errors of samples generated of RTO and MYULA with respect to MMSE estimates.}
	\label{fig:relative_err}
\end{figure}

\Cref{fig:relative_err} shows relative errors of the samples generated by RTO and MYULA to their corresponding MMSEs.
It is interesting to
note that the relative error is higher in MYULA's case. It means that the RTO posterior is more concentrated around its mean than MYULA's.
In RTO the perturbation takes place in the data space, and then is smoothed due to the use of the regularization.

\begin{figure}
		\centering
		\begin{tabular}{ccc}
			\includegraphics[width=0.3\textwidth]{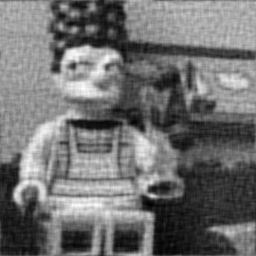}
			&
			\includegraphics[width=0.36\textwidth]{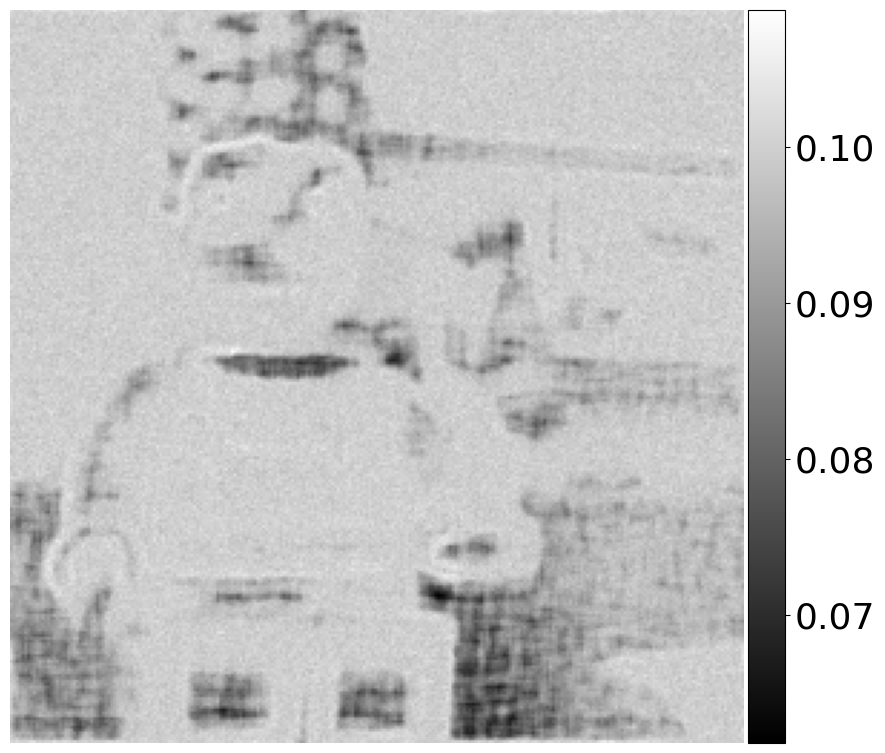}
			&
			\includegraphics[width=0.32\textwidth]{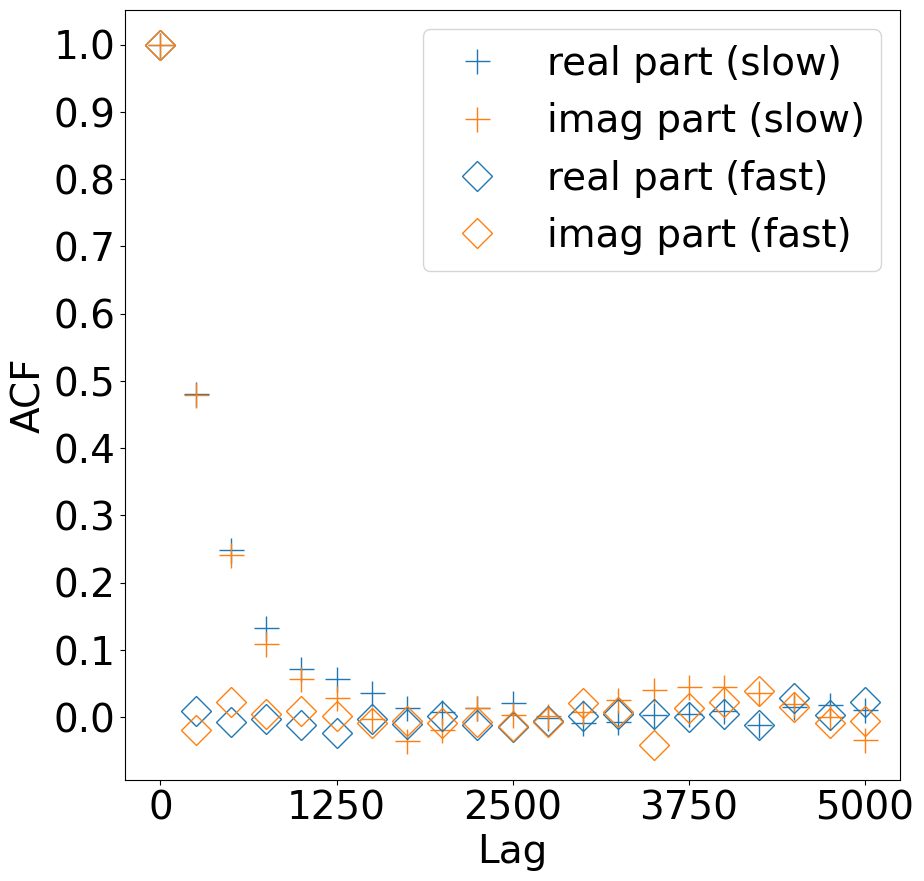}
			\\
			MMSE (PSNR=25.06/SSIM=0.66) & Standard deviation & ACFs 
		\end{tabular}
		\caption{\textit{(Deblurring)} Results generated by MYULA with $n_{pd}=10$.}
		\label{fig:myula_accuracy}
	\end{figure}

\begin{figure}
		\centering
		\begin{tabular}{cc}
			\includegraphics[width=0.3\textwidth]{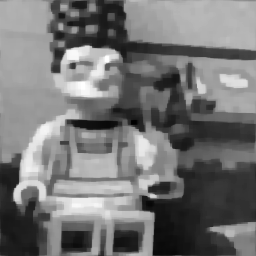}
			&
			\includegraphics[width=0.36\textwidth]{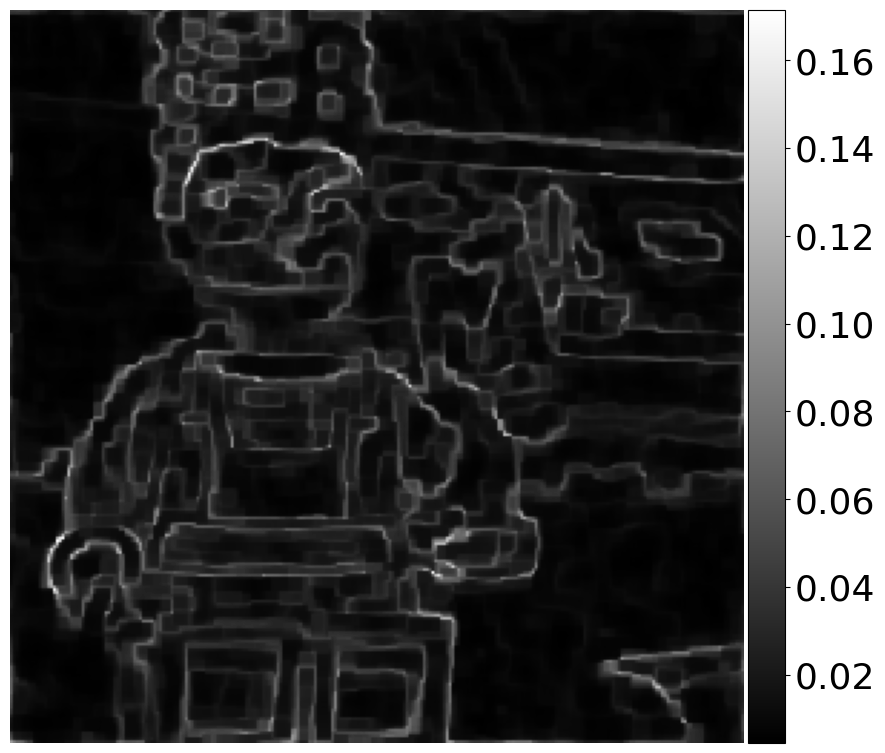}
			\\
			MMSE (PSNR=25.71/SSIM=0.79) & Standard deviation 
		\end{tabular}
		\caption{\textit{(Deblurring)} Results generated by RTO with $\texttt{tol}=10^{-2}$.}
		\label{fig:rto_accuracy}
	\end{figure}

For the computational time, it took around $6$ hours to generate 1000 RTO samples
and around $10$ hours to generate $2.5 \times 10^5$ MYULA samples after burn-in
on an Intel(R) Core(TM) i9-9980HK CPU @ 2.40GHz. The cost of generating a sample using RTO is higher than with MYULA. To reduce the computational cost,
we can decrease the accuracy when solving the optimization problems in the sampling methods,
i.e., the computation of $\operatorname{prox}_{\gamma \|\nabla . \|_{1,1}}^{\alpha_1}$ in MYULA and MAP on
the perturbed data in RTO. In \Cref{fig:myula_accuracy} we show the results produced by MYULA when setting
$n_{pd}=10$ for \texttt{Simpson}. In this case, the running time is reduced to around $3$ hours. Comparing with
the results in \Cref{fig:deblurring_tv_mmse} and \Cref{fig:deblurring_tv_std} where we set $n_{pd}=50$, we can see
that MMSE is more noisy and the uncertainty is not well discovered. In addition, the ACFs figure
shows that we need many more iterations in order to obtain uncorrelated samples. In RTO we change $\texttt{tol}$ in
ADMM from $10^{-4} $ to $10^{-2}$ and show the results in \Cref{fig:rto_accuracy}. It is clear that MMSE
is nearly identical to the one in \Cref{fig:deblurring_tv_mmse}. However, the standard deviation close to the edges are
slightly smaller than the ones in \Cref{fig:deblurring_tv_std}. It means that reasonably reducing the solution
accuracy has a minor influence on the MMSE estimate, but can lead to insufficiently explored uncertainty.

\subsection{Inpainting}\label{sec:inpainting}

\begin{figure}
\centering
	\begin{tabular}{cc}
		\includegraphics[width=0.3\textwidth]{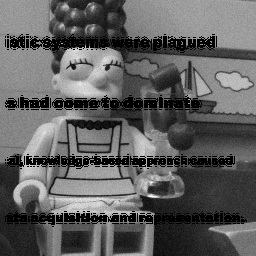} &
		\includegraphics[width=0.3\textwidth]{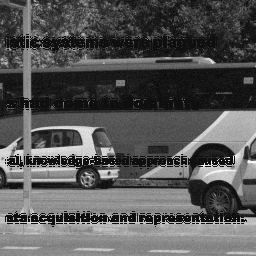}
		\\
		PSNR=16.45/SSIM=0.68 & PSNR=15.89/SSIM=0.74
	\end{tabular}
	\caption{\textit{(Inpainting)} Observations for the inpainting problem.}
	\label{fig:inp_obs}
\end{figure}

In this section, we study the inpainting inverse problem, where the forward operator $A$ indicates the
locations of the known pixel values.  
\Cref{fig:inp_obs} shows the observations $y$, where the black regions mark the information of pixels is lost.
In this test, $58549$ of $65536$ 
pixels are observed, i.e., $A\in\mathbb{R}^{58549\times65536}$. In addition, $y$ is also corrupted by additive
white Gaussian noise with zero mean and
standard deviation $\sigma=0.02$. For both methods we set the regularization parameter $\gamma =8$. To ensure good results in a reasonable amount of time, we set the stopping criteria for ADMM
in RTO as that the primal residual is smaller than $\texttt{tol} = 2\times 10^{-3}$ or the number
of iterations is larger than $500$.

\begin{figure}
	\begin{tabular}{cccc}
		& RTO & MYULA & MAP \\
		\CenteredVcell{0.3\textwidth}{\texttt{Simpson}} &
		\includegraphics[width=0.3\textwidth]{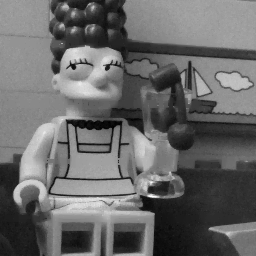} &
		\includegraphics[width=0.3\textwidth]{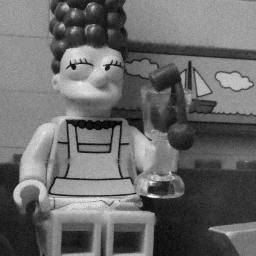} &
		\includegraphics[width=0.3\textwidth]{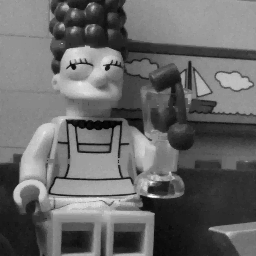}
		\\
		& \textbf{PSNR=35.01}/SSIM=0.94 & PSNR=33.85/SSIM=0.89 & PSNR=34.99/\textbf{SSIM=0.95} 
		\\
		\CenteredVcell{0.3\textwidth}{\texttt{Traffic}} &
		\includegraphics[width=0.3\textwidth]{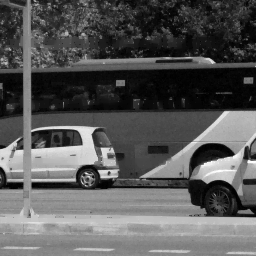} &
		\includegraphics[width=0.3\textwidth]{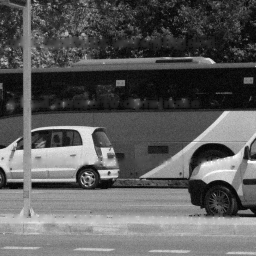} &
		\includegraphics[width=0.3\textwidth]{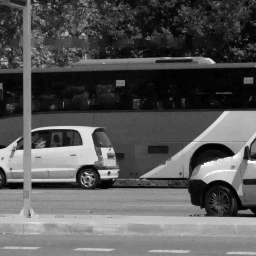} 
		\\
		& \textbf{PSNR=30.77/SSIM=0.92} & PSNR=30.44/SSIM=0.89 & PSNR=30.63/\textbf{SSIM=0.92}
	\end{tabular}
	\caption{\textit{(Inpainting)} MMSE estimates respectively obtained of RTO and MYULA together with MAP estimates.}
	\label{fig:inp_mmse}
\end{figure}

\textbf{Results and discussions:}
\Cref{fig:inp_mmse} shows MMSEs obtained by RTO and MYULA as well as MAPs. MMSEs of both methods are
very similar to the corresponding MAP estimates, and the mask regions are well filled except the regions with rich
textures, see the tree leafs in \texttt{Traffic}, where the inpainting task is the most difficult. This good
visual impression is confirmed by the high PSNR and SSIM scores. 
However, both methods tend to struggle to correctly fill the gaps as we can see on the contours of the dress
in \texttt{Simpson}. In addition, MMSEs of MYULA are slightly more noisy, and MYULA seems to have more
difficulties to fill the regions with rich textures, see the tree leafs in \texttt{Traffic}.

\begin{figure}
\centering
	\begin{tabular}{ccc} 
		\CenteredVcell{0.3\textwidth}{RTO} 
		\includegraphics[width=0.3\textwidth]{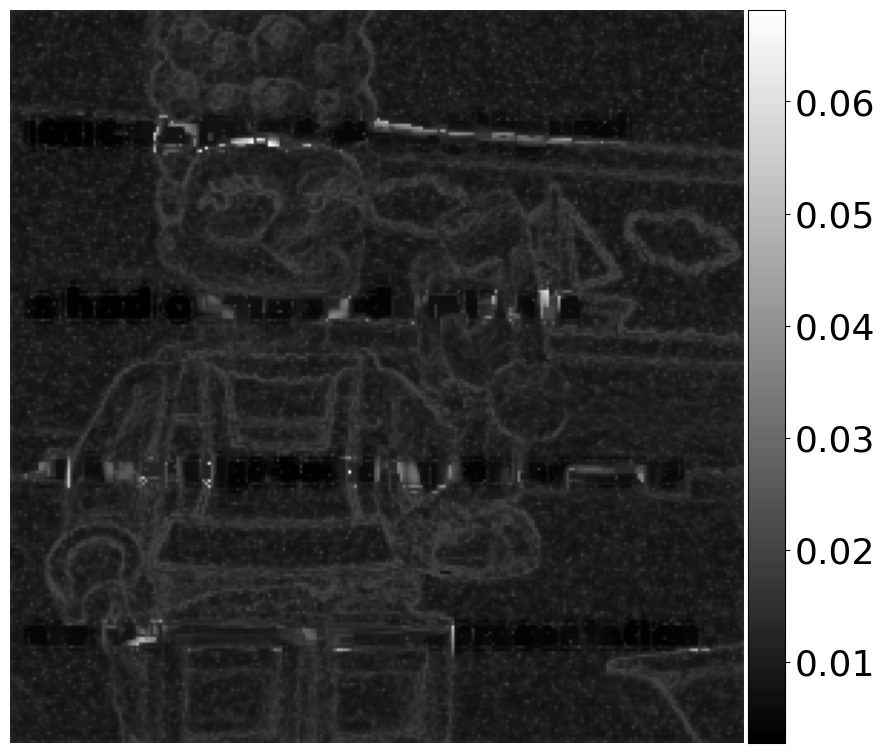} &
		\includegraphics[width=0.3\textwidth]{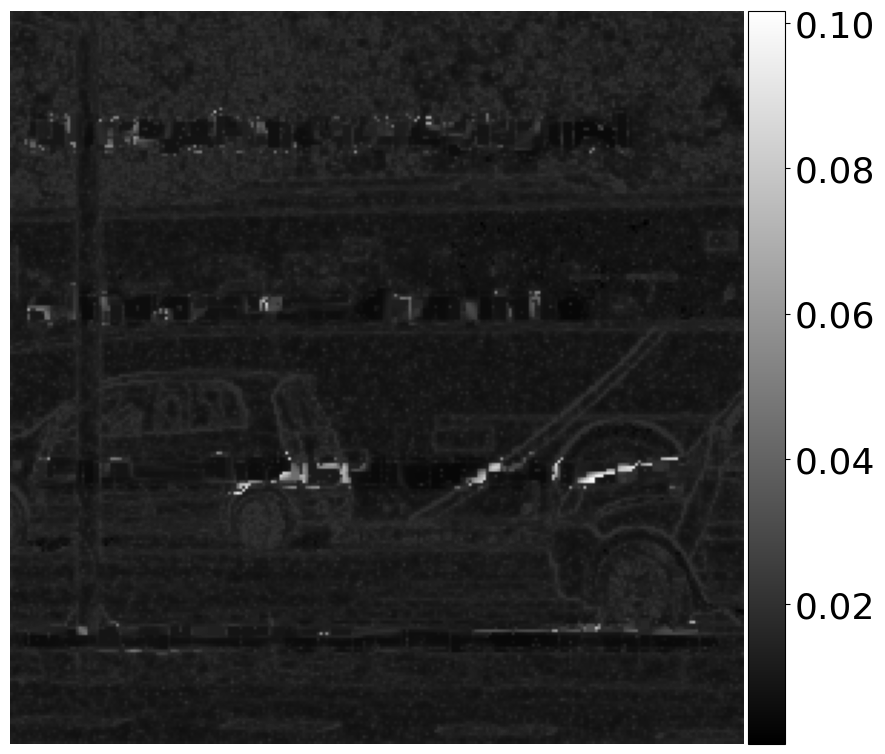} &
		\\
		\CenteredVcell{0.3\textwidth}{MYULA} 
		\includegraphics[width=0.3\textwidth]{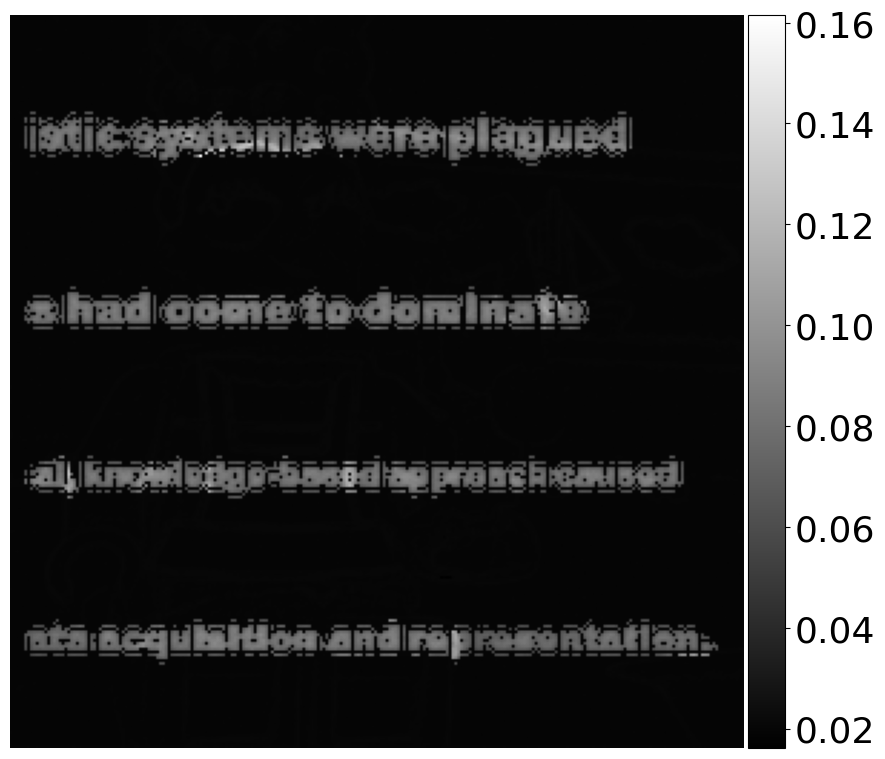} &
		\includegraphics[width=0.3\textwidth]{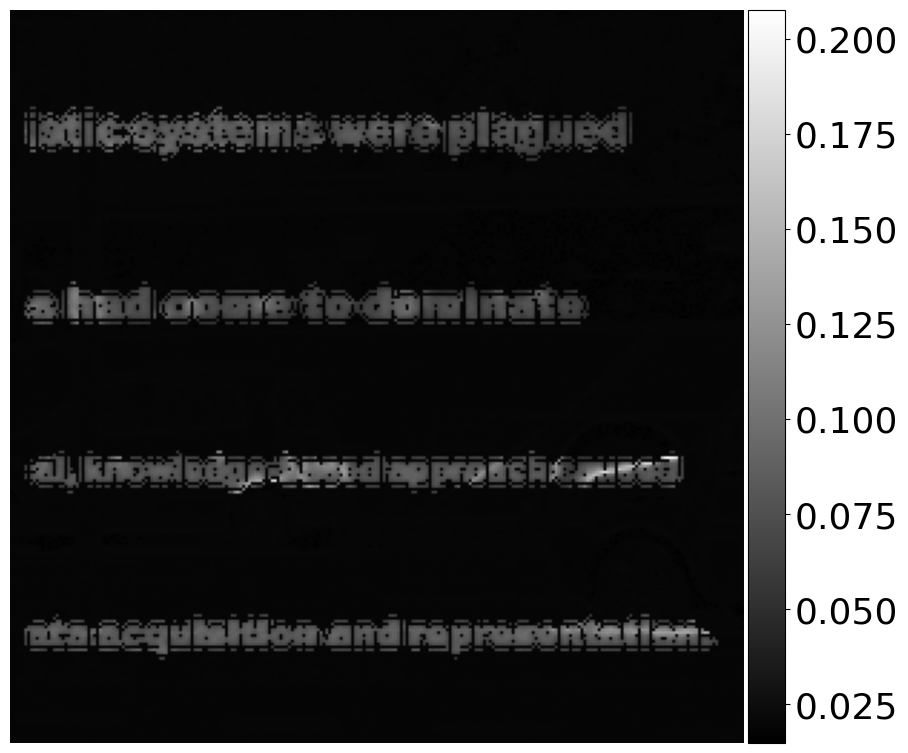}
	\end{tabular}
	\caption{\textit{(Inpainting)} The standard deviations computed with the samples generated by RTO and MYULA.}
	\label{fig:inp_std}
\end{figure}

The standard deviation maps associated with RTO and MYULA for \texttt{Simpson} and \texttt{Traffic} are displayed in \Cref{fig:inp_std}.
We can see that the standard deviations of RTO show larger uncertainties around edges, especially at the edges
where the information of pixels is lost.
It is also interesting to note that the hidden constant
areas are restored with high confidence for RTO. It means that the solutions to the perturbed optimization problems
are very similar in these regions and that the perturbations do not affect the restorations in these areas.
MYULA behaves very differently from RTO. The high uncertainties appear at the whole regions where the information
is lost. Further, 
in these regions the standard deviations from MYULA are nearly double of the RTO values.

Finally, we consider the efficiency of both methods. The experiments took around
$10$ hours both RTO and MYULA.

\begin{figure}
\centering
	\begin{tabular}{ccc}
		& \texttt{Simpson} & \texttt{Traffic}
		\\
		\CenteredVcell{0.3\textwidth}{RTO} &
		\includegraphics[width=0.3\textwidth]{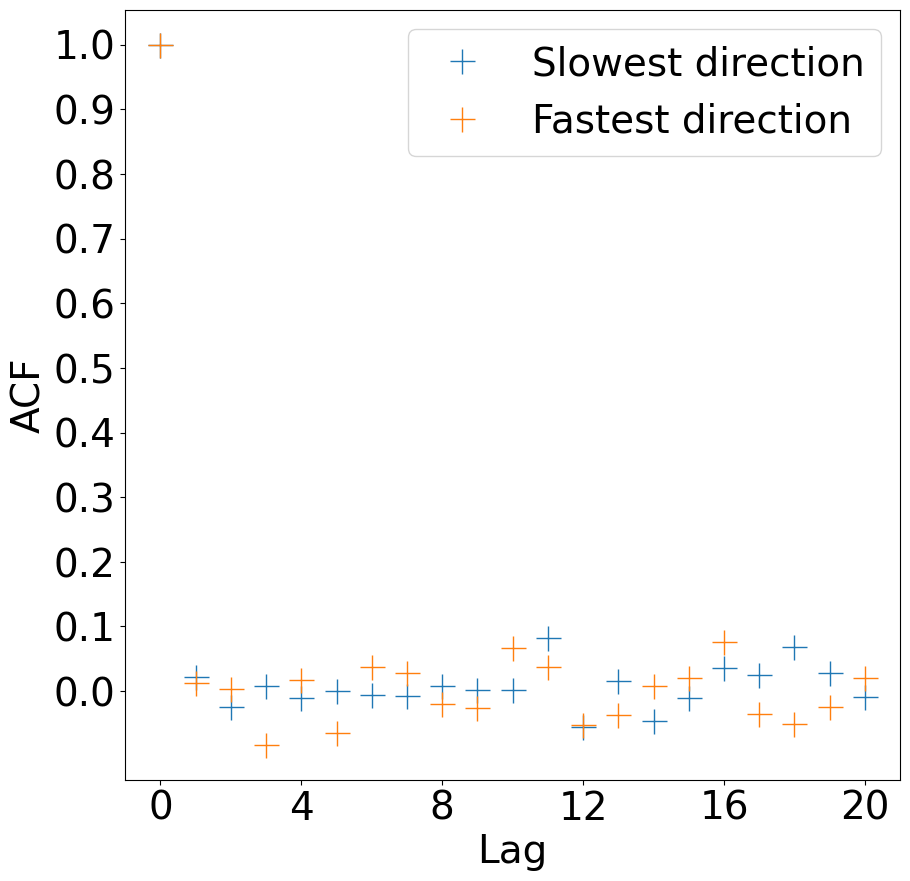} &
		\includegraphics[width=0.3\textwidth]{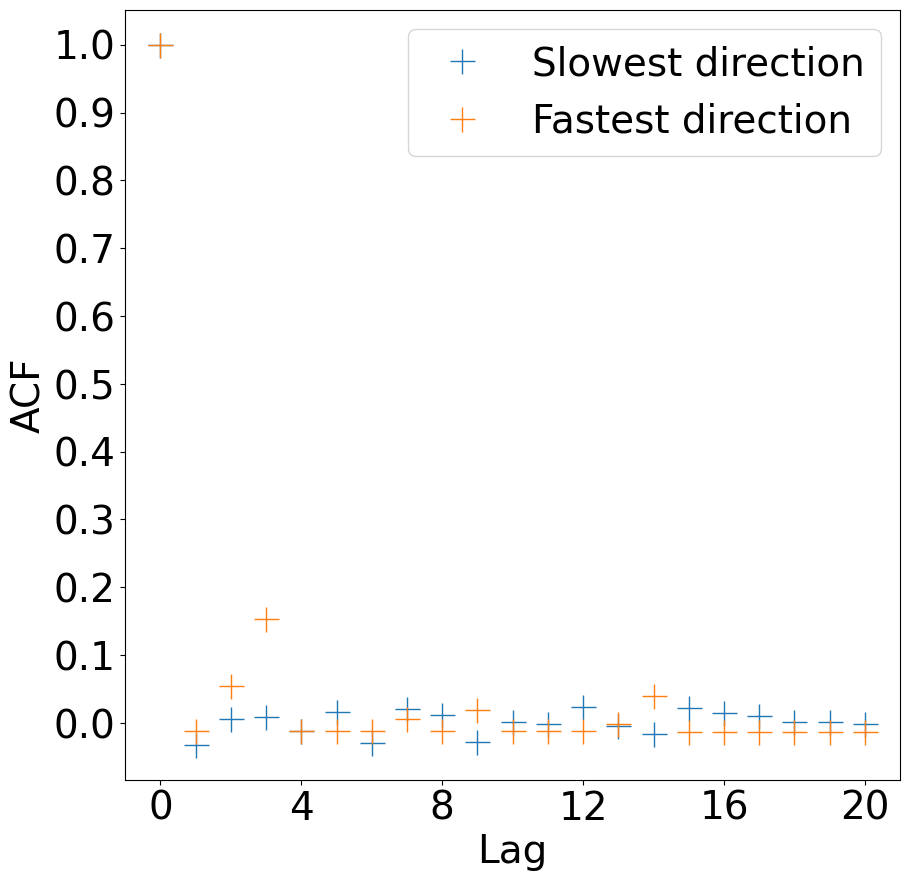}
		\\
		\CenteredVcell{0.3\textwidth}{MYULA} &
		\includegraphics[width=0.3\textwidth]{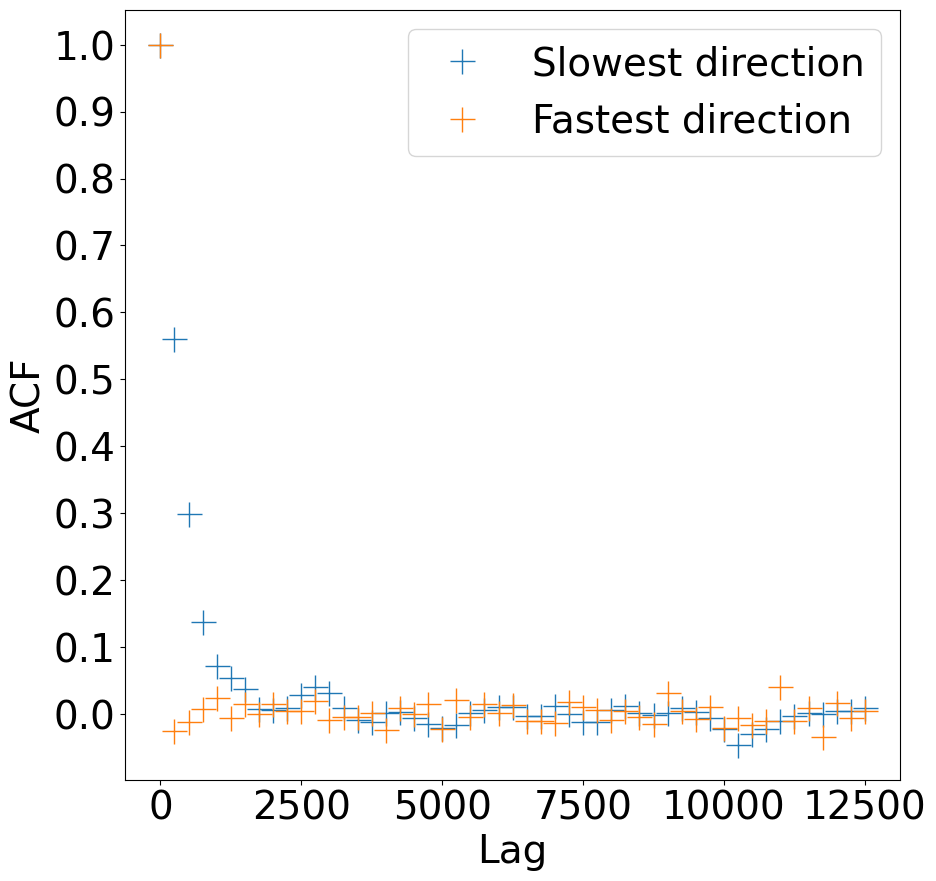} &
		\includegraphics[width=0.3\textwidth]{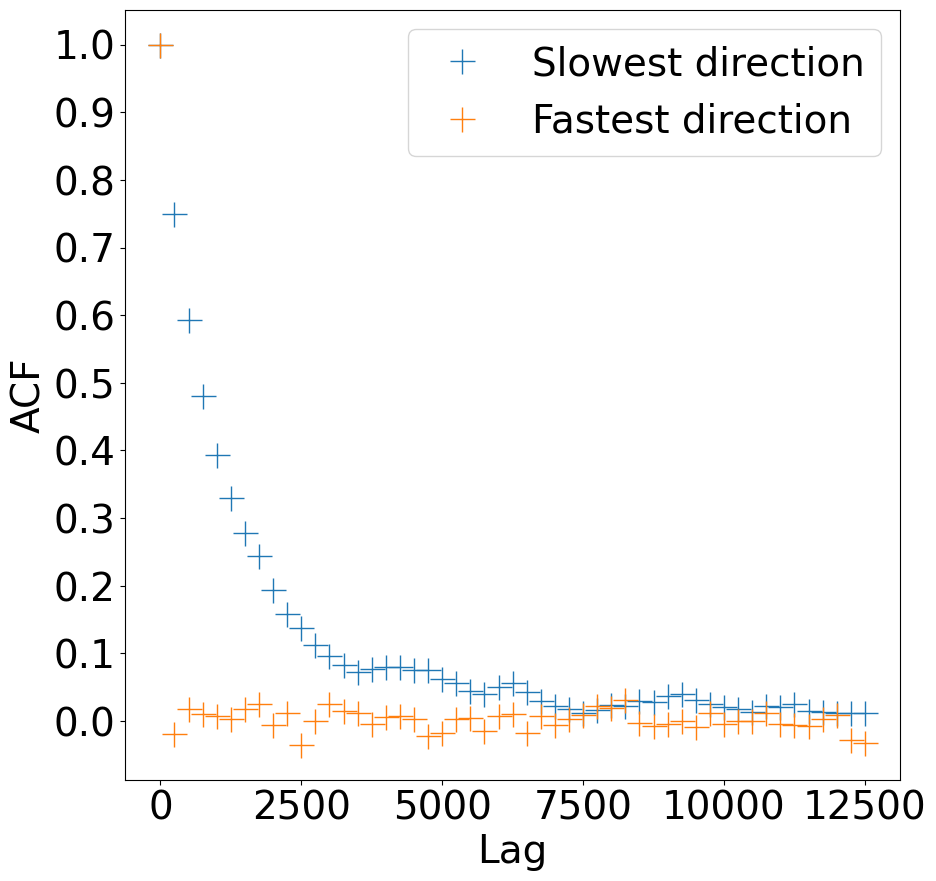}
	\end{tabular}
	\caption{\textit{(Inpainting)} ACFs from samples generated by RTO and MYULA, respectively.}
	\label{fig:inp_acf}
\end{figure}

In \Cref{fig:inp_acf} we show the evolution of the ACFs for both methods. We computed them in the pixel domain, where the forward operator $A$ is diagonal, as 
we assume that the likelihood imposes the posterior covariance shape. 
It is obvious that samples generated by RTO are independent, but samples generated with MYULA exhibit correlations. For \texttt{Traffic},
the complete Markov chain associated with MYULA needs more than $6000$ iterations in order to eventually achieve
uncorrelated samples. However, recall that generating one sample
with RTO is approximately 250 times more expensive than with MYULA.

\subsection{Automatic parameter selection}\label{sec:gibbs_rto}

In \Cref{sec:deblurring,sec:inpainting} we choose the regularization parameter $\gamma$ such that it gives the best results in terms of PSNR and SSIM, which requires many more experiments in order to determine $\gamma$. Furthermore, we assume that we have access to the noise level $\sigma$ that is not always known.
Being able to automatically select the parameters in a robust fashion is a matter of prime importance.
As explained in \Cref{sec:theory_comp}, Langevin methods like MYULA do not allow us to perform online
parameter selection, but RTO can be plugged into an augmented hierarchical model in order to automatically select
$\sigma^2$ and $\gamma$ without any
additional computational cost \cite{Everink_2023,everink2023bayesian}, see \Cref{alg:augmented_rto}.
It turns RTO into a nearly parameter-free method. To ensure efficient sampling for the conditional distributions
with respect to $\sigma^2$ and $\gamma$, we are limited to $\Gamma$-priors for $\lambda =1/\sigma^2$ and $\gamma$,
and only for some specific choices of $g$. For more detailed discussions, we refer to \cite{everink2023bayesian}.

\begin{figure}
	\centering
	\begin{tabular}{cc}
		MMSE & Standard deviation
		\\
		\includegraphics[width=0.3\textwidth]{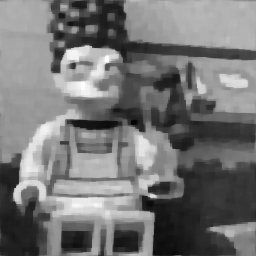} &
		\includegraphics[width=0.37\textwidth]{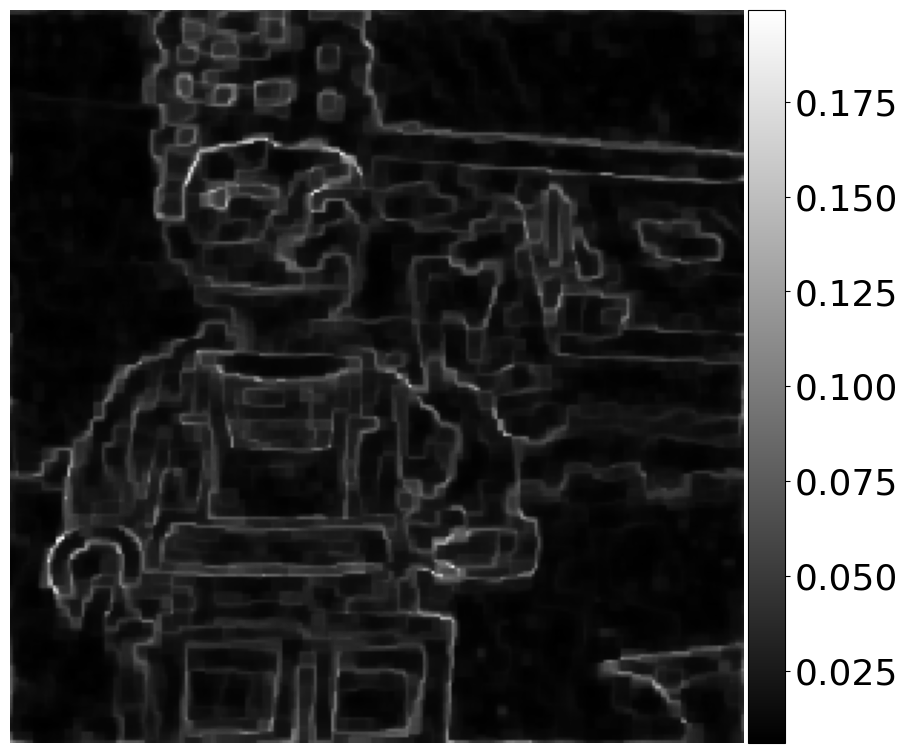} 
		\\
		PSNR=26.00/SSIM=0.79 &
	\end{tabular}
	\caption{\textit{(Hierarchical Gibbs sampler)} MMSE and marginal posterior standard deviations computed using the hierarchical Gibbs sampler
	presented in Algorithm \ref{alg:augmented_rto} for image deblurring problem defined in Section \ref{sec:deblurring}.}
	\label{fig:gibbs_mmse_std}
\end{figure}

\begin{figure}
	\centering
	\begin{tabular}{c||cc}
		\includegraphics[width=0.3\textwidth]{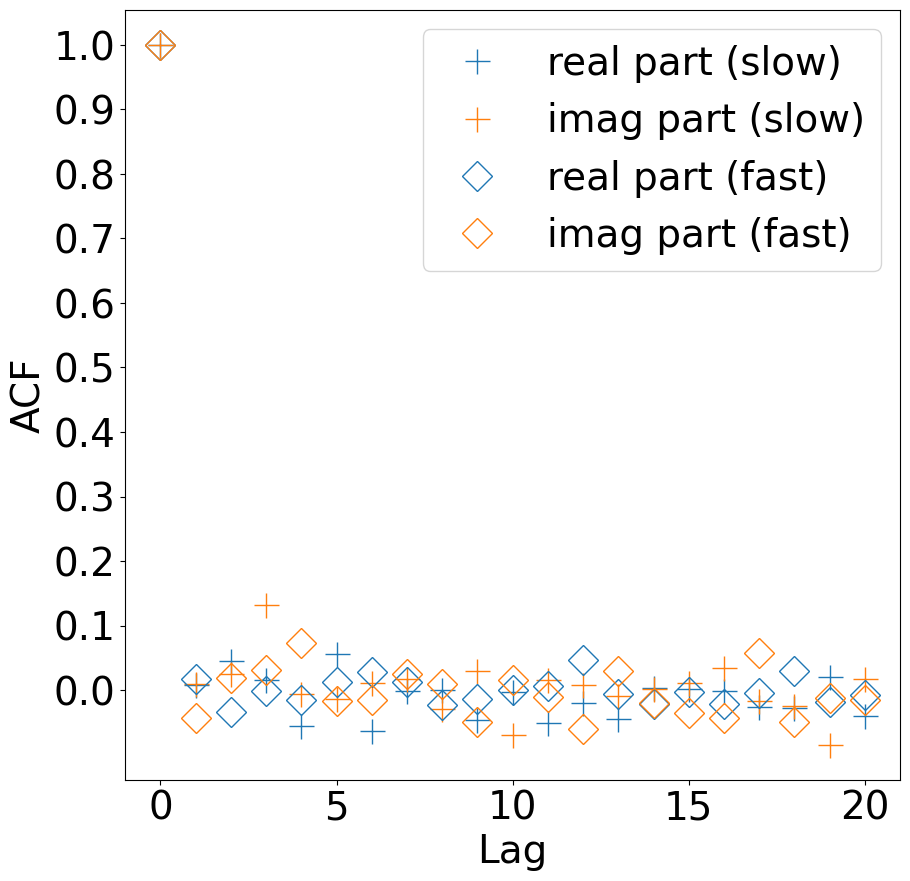}
		&
		\includegraphics[width=0.3\textwidth]{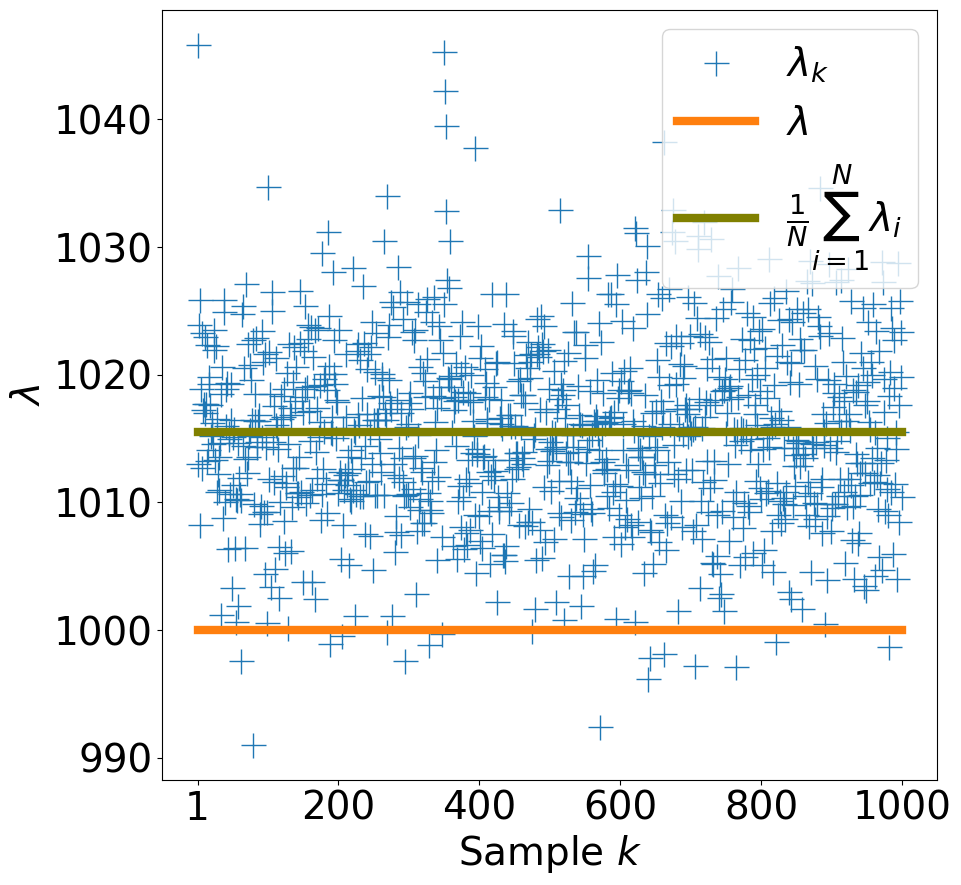}
		&
		\includegraphics[width=0.3\textwidth]{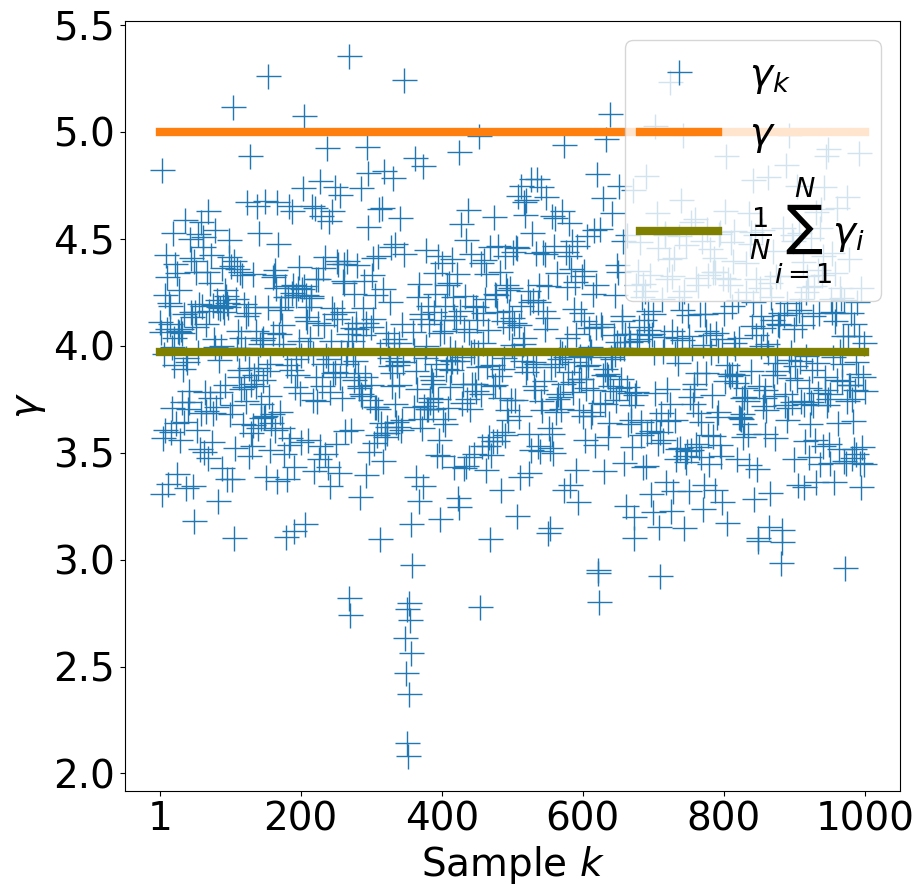}	
		\\
		ACFs & Trace $\lambda_k$ & Trace $\gamma_k$
	\end{tabular}
	\caption{\textit{(Hierarchical Gibbs sampler)} Analysis of the sample correlation of the augmented Gibbs sampler.
	Left: ACFs of the image samples. Middle and right: traces of the noise level and regularization parameters.}
	\label{fig:gibbs_acf}
\end{figure}

In \Cref{fig:gibbs_mmse_std} we show the results from the same deblurring problem as in \Cref{sec:deblurring} but
with slightly different $g(x)$ comparing with \eqref{eq:gnu}. Here we use the non-negativity constraint instead of
the box constraint, i.e., $g(x) = \gamma\|\nabla x \|_{1, 1} + i_{\mathbb{R}^+}(x) $, since in this case the
conditional distribution $\pi_{\bbgamma|\mathbbm{x},\mathbbm{y}}(\gamma)$ can be efficiently sampled. Comparing the
results shown in \Cref{fig:gibbs_mmse_std} and in \Cref{fig:deblurring_tv_mmse,fig:deblurring_tv_std}, we can see
that they are very similar visually as well as quantitatively. In \Cref{tab:gibbs_parameter}, we list the means and
the standard deviations of $\lambda$ and $\gamma$ obtained from the hierarchical Gibbs sampler. It is clear that
the estimates for both $\gamma$ and $\lambda$ are comparable with the ones used in \Cref{sec:deblurring}. In the
end, \Cref{fig:gibbs_acf} shows the ACFs of the image samples and trace plots of $\lambda$ and $\gamma$. Due to the
Gibbs sampler, we cannot expect independent samples, but the trace plots show very good mixing properties.
It indicates that the samples produced by our augmented model are practically uncorrelated \cite{bardsley2018computational,robert_monte_2004}.

\begin{table}[t]
	\centering
	\begin{tabular}{|c|c|c|}
		\hline
		& Obtained from hierarchical Gibbs sampler & Used in RTO in \Cref{sec:deblurring} \\
  & \textbf{mean} (standard deviation) & \\
		\hline
		$\gamma$ & \textbf{3.97} (0.44) & 5 \\
		\hline
		$\lambda$ & \textbf{1015.49} (7.19) & 1000 \\
		\hline
	\end{tabular}
	\caption{\textit{(Hierarchical Gibbs sampler)}  Empirical mean and standard deviation of the noise level and regularization parameters $\lambda$ and $\gamma$.}
	\label{tab:gibbs_parameter}
\end{table}

\section{Conclusion}\label{sec:conclusion}

We compared two classes of sampling methods for solve inverse problems in imaging,
RTO and the Langevin method MYULA, and highlighted their main conceptual and theoretical differences. 

RTO is derived from the sensitivity analysis framework and samples from a target
distribution by solving perturbed optimization problems where the perturbation
occurs in the data space. The RTO target density can have non-zero probability
mass on subsets of measure zero. However, it is not anchored in the Bayesian
framework as the target density corresponds to an implicit prior that depends on
the observed data. In addition, RTO can be incorporated into a hierarchical model
in order to perform automatic parameter selection. The main limitation of RTO is
that it has only been characterized for the posteriors with Gaussian likelihoods
and polyhedral hypograph log-prior. Its efficiency also depends on
the ability to efficiently solve perturbed MAP optimization problems.
In contrast, MYULA is firmly rooted in the
Bayesian framework and is applicable to a broader range of posteriors.
Generating a single sample with MYULA is rather cheap. Although
it samples the posterior density approximately, the distribution behind samples
can be characterized with respect to the posterior density.
Through two classical imaging inverse problems:
deblurring and inpainting, we compared RTO and MYULA numerically with particular
attention to computational cost. Both methods produced accurate results for
deblurring, but MYULA struggled with severely ill-posed problems like inpainting.
Additionally, while RTO concentrates the sample mass around the MMSE estimate,
MYULA results in a more dispersed distribution.

One future research direction is to extend RTO to more general posteriors. It would
be valuable to explore other noise models, such as Poisson noise as in \cite{bardsley2020mcmc},
and investigate how we can characterize the RTO distribution both theoretically
and practically. In addition, motivated by the work in \cite{laumont2022bayesian},
it would be also interesting to extend RTO to data-driven regularization
\cite{pesquet2021learning,hurault2022proximal,mukherjee2024data}, particularly to
generative models \cite{rezende2015variational,kingma2019introduction}.

\bibliographystyle{unsrt}
\bibliography{main.bib}

\end{document}